\documentclass[fleqn,usenatbib]{mnras}

\usepackage{newtxtext,newtxmath}
\usepackage[T1]{fontenc}
\usepackage{ae,aecompl}

\usepackage{graphicx}	% Including figure files
\usepackage{dblfloatfix}
\usepackage{amsmath}	% Advanced maths commands
\usepackage{threeparttable}
\usepackage{subcaption}
\usepackage{caption}
\usepackage{color}
\usepackage[dvipsnames]{xcolor}
\usepackage[normalem]{ulem}
\usepackage{booktabs}
\newcommand{\fermi}{\textit{Fermi}-{\rm LAT}}
\newcommand{\planck}{\textit{Planck}}
\newcommand{\gray}{$\gamma$-ray}
\newcommand{\grays}{$\gamma$-rays}

\newcommand{\msun}{\mbox{$M_\odot$}}
\def\deg{\hbox{$^\circ$}}

\title[\fermi\ detection around the W3 complex]{Extended GeV \gray\ emission around the massive star forming region of the W3 complex}
\author[Wu et al.]{Qi-Hang Wu,$^{1}$
Xiao-Na Sun,$^{1}$\thanks{E-mail:xiaonasun@gxu.edu.cn}
Rui-Zhi Yang,$^{2,3,4}$
Ting-Ting Ge,$^{5,6}$
Yun-Feng Liang$^{1}$
\newauthor{and En-Wei Liang$^{1}$}
\\
$^{1}$Guangxi Key Laboratory for Relativistic Astrophysics, School of Physical Science and Technology, Guangxi University, Nanning 530004, China\\
$^{2}$Department of Astronomy, School of Physical Sciences, University of Science and Technology of China, Hefei, Anhui 230026, China\\
$^{3}$CAS Key Laboratory for Research in Galaxies and Cosmology, University of Science and Technology of China, Hefei, Anhui 230026, China\\
$^{4}$School of Astronomy and Space Science, University of Science and Technology of China, Hefei, Anhui 230026, China\\
$^{5}$School of Physics and Astronomy, Sun Yat-sen University, Zhuhai 519082, China\\
$^{6}$CSST Science Center for the Guangdong-Hong Kong-Macau Greater Bay Area, Sun Yat-Sen University, Zhuhai 519082, China\\
}

\pubyear{2023}
\begin{document}
\label{firstpage}
\pagerange{\pageref{firstpage}--\pageref{lastpage}}

\maketitle

\begin{abstract}
We analyze the GeV \gray\ emission from the W3 complex using about 14 years of Pass 8 data recorded by the $\it Fermi$ Large Area Telescope (\fermi).
We resolve the \gray\ emissions around W3 into two components: an elliptical Gaussian overlapping with the molecular gas and a point-like source near the cluster W3 Main.
The pion-bump feature of SED for the elliptical Gaussian together with the better fitting result of pion decay model favor the hadronic origin.
We further argue that the cosmic rays (CRs) could originate from the interactions between cluster winds and the shock produced by the SNR HB3.
The point-like source positionally coincident with the star cluster W3 Main indicates it may be directly powered by near clusters, while its fainter \gray\ emissions below 10 GeV is possibly due to the shelter from dense gas making the low-energy CRs incapable of penetrating the dense materials.
Meanwhile, we cannot rule out that the \gray\ emissions originate from the interaction of accelerated protons in SNR with the ambient gas.
\end{abstract}

% Select between one and six entries from the list of approved keywords.

%%%%%%%%%%%%%%%%%%%%%%%%%%%%%%%%%%%%%%%%%%%%%%%%%%

%%%%%%%%%%%%%%%%% BODY OF PAPER %%%%%%%%%%%%%%%%%%

%%%%%%%%%%%%%%%%% introduction %%%%%%%%%%%%%%%%%%
\section{Introduction}

The issue on the origin of CRs has been existing for many years.
In the CR community, there is a consensus that CRs of energy below $\sim10^{15}$eV (called "knee") are produced in the Milky Way \citep{Aloisio2007}.
Supernova remnants (SNRs) have been considered as the main acceleration sites of Galactic CRs for several decades since the diffusive acceleration by supernova shock waves can accelerate particles to very high energies \citep{Lagage1983, Bell2013}.
Young massive stellar clusters (YMCs) have also been supposed to be potential sites of CR acceleration \citep{Parizot2004A&A...424..747P}.
YMCs typically host a large number of massive stars which can drive high-speed stellar winds almost sustaining the lifetime \citep{Portegies2010ARA&A..48..431P}.
\cite{Aharonian2019NatAs...3..561A} detected hard spectra of gamma-rays and CRs across Cygnus Cocoon and Westerlund 1, and derived 1/r radial profile of CR energy density to characterise the continuously central injection of YMCs.
Also, a number of GeV-TeV \gray\ sources are found in the direction of various YMCs, e.g.,
Westerlund 2 \citep{Yang2018A&A...611A..77Y}, NGC 3603 \citep{Yang2017A&A...600A.107Y}, 30 Dor C \citep{30DorC2015Sci...347..406H}, RSGC 1 \citep{sunRSGC1}, W40 \citep{sunw40}, Mc20 \citep{sun22}, NGC 6618 \citep{Liu2022}, and Carina Nebula Complex \citep{Ge2022}.
Some multi-wavelength simulations and theory calculations also indicate that YMCs have the capability of to explain the problem of the origin of CRs from SNRs to some extent, such as the maximum particle energy and isotopic composition \citep{Gupta2018MNRAS.479.5220G,Gupta2020MNRAS.493.3159G}.
Nevertheless, it must be noted that there are hardly any clear-cut identification of particle acceleration by YMCs.
Studies of Cygnus cocoon \citep{Astiasarain2023A&A...671A..47A} and Westerlund 1 \citep{Bhadra10.1093/mnras/stac023} shows that the population can be hadronic or leptonic.
The 1/r profile which was derived by neglecting some crucial aspects (e.g., advection of CRs, various acceleration sites and radiative loss, etc.) may not reflect the true radial profile, and alternative scenarios such as CRs injected in the wind termination shock region or discrete multiple SN injections are able to yield a 1/r profile \citep{Bhadra10.1093/mnras/stac023}.
Yet, the growing number of \gray\ source towards YMCs as well as characterisations found in YMCs are helpful and to some extent in favor of YMCs from which the contribution to CRs cannot be rule out.

W3 is one of the most active and nearest massive star forming regions located in the Perseus Arm of the outer galaxy \citep{Reid2016}.
The giant molecular cloud (GMC) W3 was first discovered through the radio continuum emission \citep{Westerhout1958} and the total mass is $\sim$4 $\times\ 10^5\msun$ \citep{Polychroni2012MNRAS.422.2992P}.
Its kinematic distance to the Sun measured by  the Galactic rotation curve is $\sim$4.2 kpc \citep{Russeil2003A&A...397..133R}, which is significantly different from the value of about 1.9 - 2.4 kpc estimated by trigonometric and spectrophotometric methods \citep{Routledge1991,Hachisuka2006,Xu2006Sci...311...54X,Navarete2011}.
This region of the Perseus arm does not follow the Galactic rotation curve as discussed by \cite{Navarete2011}.
The discrepancy between the kinematic distance to W3, which is roughly twice the distance for non-kinematic methods, may be attributed to local motions of the gas deviating from the Galactic rotation.
Velocity anomalies and the peculiar motion of the Perseus arm have been detected by \cite{Brand1993A&A...275...67B}.
Also, \cite{Russeil2003A&A...397..133R} noted a velocity discrepancy of the Perseus arms of 21 km s$^{-1}$, which is significant with respect to typical values measured for other spiral arms ($\sim$3 km s$^{-1}$).
Adjacent to W3 in the east direction is the W4 $\ion{H}{II}$ region ionized by IC 1805 \citep{Massey1995ApJ...454..151M} along the Galactic Plane.
In the boundary between W3 and W4 there is a high density layer (HDL), above $\sim10^{22}$ cm$^{-2}$, containing about half of the total mass of the cloud \citep{Polychroni2012MNRAS.422.2992P}.
Bright $^{12}$CO(J = 1–0) line emission near $-43$ km s$^{-1}$ is observed in the vicinity of the W3, especially near the HDL.
\cite{Yamada2021} argued that the radial velocity of gas surrounding the GMC W3 is around $-53 \sim -37$ $\rm km\ s^{-1}$ in which the velocity range of $-53 \sim -41$ $\rm km\ s^{-1}$ is connected to W3(OH) and the velocity range of $-46 \sim -37$ $\rm km\ s^{-1}$ is associated with W3 Main.
A sequence of star forming sub-regions lie along the border of the W3 cloud, such as W3 Main, IC 1795, and W3(OH).
The diffuse $\ion{H}{II}$ region IC 1795 is located between YMCs W3 Main and W3(OH) \citep{Mathys1989}.
It has been suggested that IC 1795 triggered the formation of other star forming regions in a hierarchical progression \citep{Oey2005}.
\cite{Román-Zúñiga2015} argued that IC 1795 formed first, about 3–5 Myr ago \citep{Oey2005}, followed by the W3 Main cluster located to its west edge, and the W3(OH) to the east, both with ages of 2–3 Myr according to spectroscopic studies \citep{Navarete2011,Bik2012}.
The GMC harbours totally $\sim$100 OB stars of which the O-type population stars concentrate in W3 Main and IC 1795 and the B-type stars are not confined to the HDL, giving a hint that the star formation in the W3 complex began spontaneously and is earlier than the age of the clusters \citep{Kiminki2015ApJ...813...42K}.
The Chandra study by \cite{Feigelson2008} also indicated that the clusters in W3 extend widely and are highly structured and the sources therein are located at relatively large distances from the dynamical centers.

Adjacent to the northwest of the W3 complex is a well-known middle-aged SNR HB3 (G 132.7+1.3) with a diameter of $\sim$1.3$\deg$ traced by radio data \citep{Routledge1991, Green2014}.
The $^{12}$CO(J = 1–0) line emission around W3 is partly surrounded by a region of enhanced radio continuum emission from HB3, indicating that there exist interactions between HB3 and gas from the W3 complex \citep{Routledge1991,Zhou2016}.
The distance to the SNR is therefore considered to be the same as that of W3 \citep{Zhou2016}.
The age was estimated to be $\sim 1.95 \times 10^4$ years based on X-ray data \citep{Lazendic2006}.
A pulsar with $\tau_{\rm c}$$\sim$13 Myr detected by \cite{Lorimer1998} is close to the SNR's boundary but it seems to have no correlation with the remnant.

Using about 5.5-year \fermi\ data, \cite{Katagiri2016} modelled the \gray\ emissions of W3 complex and SNR HB3 as CO template and an uniform disk which is adopted by Fermi collaboration in 4FGL catalog, and argued these \gray\ emissions have the same origin which is the interactions between the hadrons accelerated by the SNR and ambient gas.
And the pion-bump feature of W3 was firstly detected by \cite{Abdollahi2022ApJ...933..204A}.
The contribution of YMCs may play an important role in the Galactic CRs \citep{Aharonian2019NatAs...3..561A,Peron2024NatAs.tmp...10P}.
Yet the potential contribution to gamma rays and CRs of YMCs in W3 complex is not considered in \cite{Katagiri2016}.
We conduct a detailed analysis in this region taking advantage of nearly 14 years of \fermi\ data and considering the impact of star clusters.

The paper is organized as follows. 
In Sect.\ref{sec:data_analysis}, we present the data set and the results of the data analyses.
The gas distributions are derived in Sect.\ref{sec:Gas}.
And we investigate the possible origin of the \gray\ emissions in Sect.\ref{sec:origin}.
The discussion and conclusion are presented in Sect.\ref{sec:conclusion}. 

%%%%%%%%%%%%%%%%% data reduction and analysis %%%%%%%%%%%%%%%%%%
\section{\fermi\ data analysis}
\label{sec:data_analysis}

$Fermi$ Gamma-Ray Space Telescope was launched on 2008 June 11, its main instrument the Large Area Telescope (LAT) operates in the \gray\ energy band from $\sim$20 MeV to >300 GeV. The LAT has a larger field of view ($\sim$2.4 sr), a larger effective area ($\sim$8000 cm$^2$ for >1 GeV on-axis), improved point-spread function (PSF) and sensitivity compared to previous high-energy \gray\ telescopes \citep{Abdo2009,Ackermann2012ApJS..203....4A}.

We select the \fermi\ Pass 8 data toward the W3 region from August 4, 2008 (MET 239557417) until July 3, 2022 (MET 678583830), and use the standard LAT analysis software package $\it v11r5p3$ \footnote{\url{https://fermi.gsfc.nasa.gov/ssc/data/analysis/software/}}. 
A 14\deg\ $\times$ 14\deg\ square region centered at the position of W3 (R.A. = 35.62$\deg$, Dec. = 61.94$\deg$) is taken as the region of interest (ROI).
The instrument response functions (IRFs)  P8R3\_SOURCE\_V3 are adopted for SOURCE events (evclass = 128).
We also apply the recommended expression $\rm (DATA\_QUAL > 0) \&\& (LAT\_CONFIG == 1)$ to pick out the good time intervals (GTIs) based on the information provided in the spacecraft file.
In order to reduce \gray\ contamination from the Earth's albedo, only events with zenith angles less than 90$\deg$ are included in the analysis.
The source model generated using the script {\it make4FGLxml.py}\footnote{\url{https://fermi.gsfc.nasa.gov/ssc/data/analysis/user/}} is based on the 4FGL-DR3 source catalog \citep{4FGL,4FGL-DR3}.
It consists of all spectral and spatial parameters of the 4FGL sources whose positions are within a radius of 20\deg\ centered at W3 as well as the Galactic diffuse emission {\it gll\_iem\_v07.fits} and the isotropic emission {\it iso\_P8R3\_SOURCE\_V3\_v1.txt}\footnote{\url{https://fermi.gsfc.nasa.gov/ssc/data/access/lat/BackgroundModels.html}}.
We use PSF type events to perform the joint likelihood analysis and use the optimizer NEWMINUIT.
The binned likelihood analysis is performed by {\it gtlike} tool.
We firstly preform the fitting using the model produced by Fermi Collaboration, then limit the number of free parameters to less than 15, i.e., we only free the spectral parameters of two sources, W3 and HB3, and the normalization parameters of sources within 5\deg\ from the ROI center as well as the two diffuse background components.
And we account for the effect of energy dispersion by using \textit{edisp\_bins}=-3.
We note that there are two extended sources: 4FGL J0222.4+6156e and 4FGL J0221.4+6241e are associated with W3 and HB3, respectively.

\subsection{Morphological Analysis}
\label{model}

\begin{figure}
    \centering
    \includegraphics[scale=0.35]{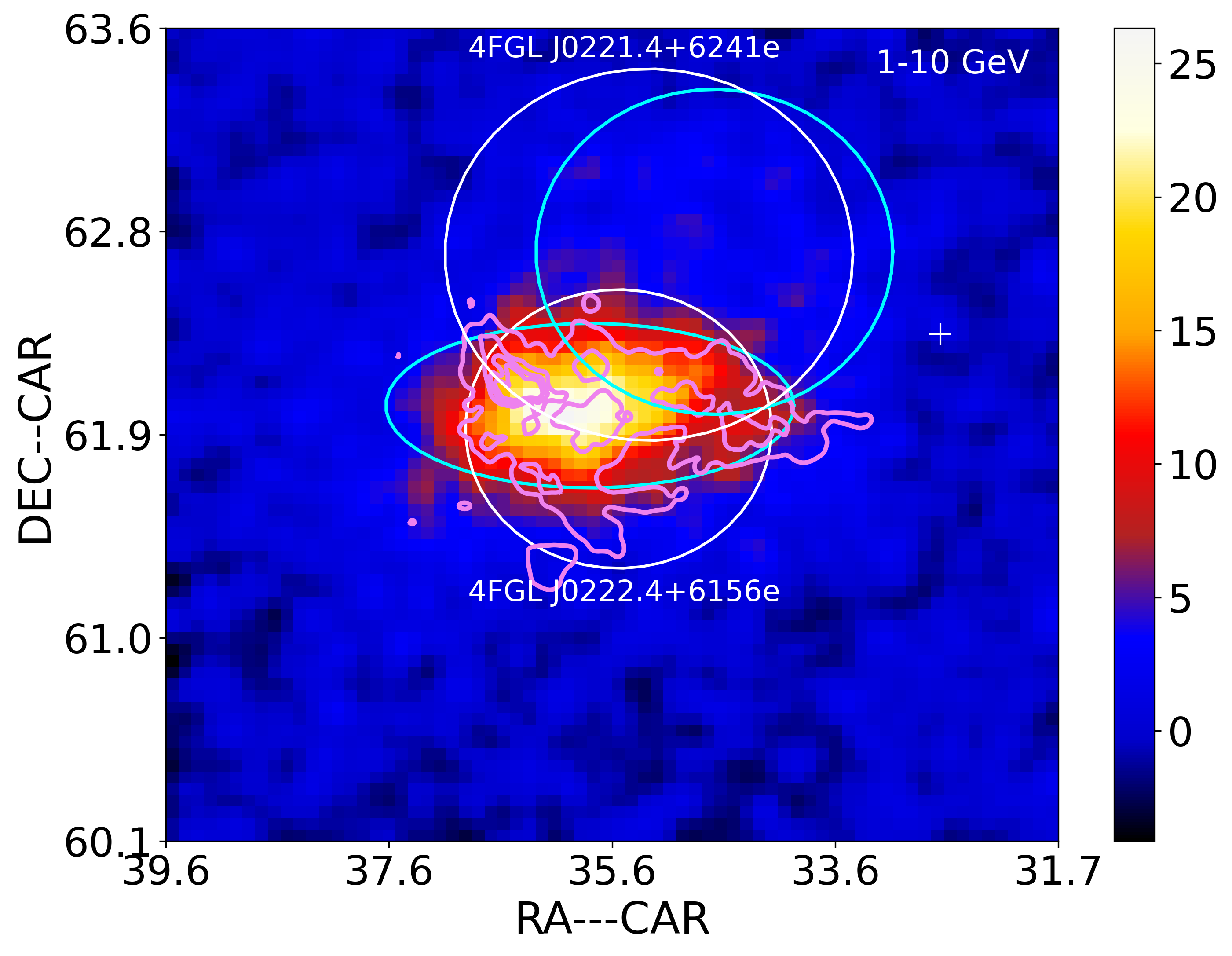}
    \includegraphics[scale=0.35]{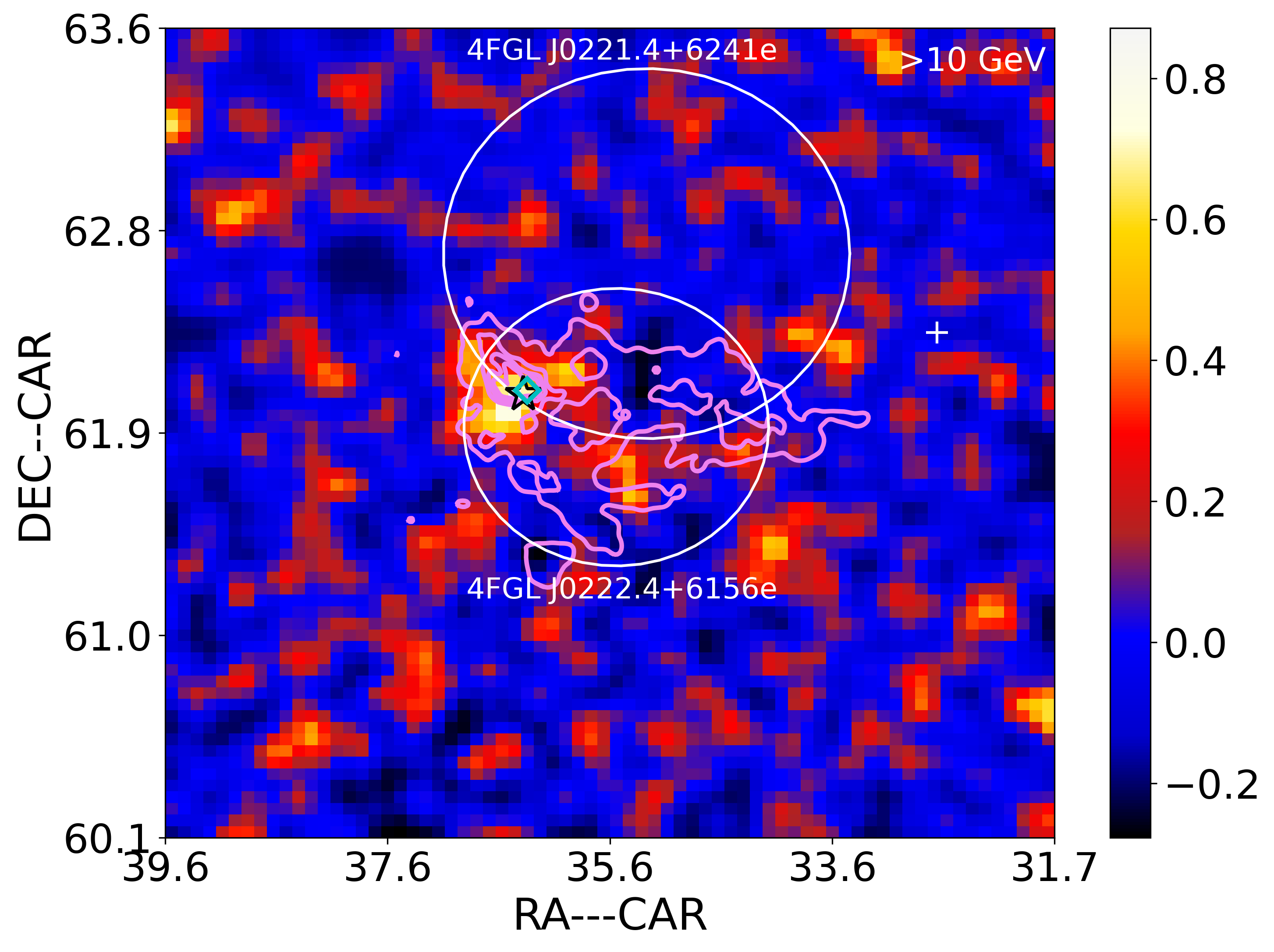}
    \caption{\fermi\ residual maps in units of counts with pixel size of $0.05^\circ$ in the $3.5^\circ \times 3.5^\circ$ region around W3 after subtracting diffuse emissions and all 4FGL sources except 4FGL J0222.4+6156e and 4FGL J0221.4+6241e which are associated with W3 and HB3 (white circles).
    The white plus indicates the point source 4FGL J0211.5+6219.
    The violet contours show the distribution of CO gas of MWISP (See details in Sect.\ref{sec:Gas}).
    The upper panel shows the residual map from 1 GeV to 10 GeV.
    The cyan ellipse and circle show the best-fit elliptical Gaussian modeling W3 complex and the optimized uniform disk modeling HB3 in the 1-10 GeV band, respectively.
    The lower panel shows the residual map above 10 GeV.
    The black asterisk denotes the peak of the residual map at which we add a new point-like source in our subsequent analysis.
    The cyan diamond near the asterisk shows the location of YMC W3 Main.
    }
    \label{fig:resmap}
\end{figure}

We split the data into four event types with associated PSF0, PSF1, PSF2, and PSF3, respectively, to avoid diluting high-quality events (PSF3) with
poorly localized ones (PSF0) to perform a joint likelihood fit for morphology analysis.
Attractively, in the analysis we find apparently energy-dependent phenomenon of \gray\ emissions.
We derive the residual maps of W3 and HB3 in 1-2, 2-5, 5-10, >10 GeV energy bands, find the morphologies of \grays\ are extended below 10 GeV and point-like above 10 GeV, and the peaks of \gray\ emissions shift from west below 10 GeV to east above 10 GeV.
Thus, as shown in Fig.\ref{fig:resmap}, we generate the \gray\ residual maps in the $3.5\deg \times 3.5\deg$ region around W3 in the 1-10 GeV and 10-300 GeV energy bands by subtracting the 4FGL J0222.4+6156e associated with W3 and 4FGL J0221.4+6241e related to HB3 from the background.
It is apparent that as the photon energy goes higher the \gray\ morphology varies from diffuse to point-like, and the emission peak shifts from west to east.
To make clear the confusion around W3, we apply specific likelihood ratio tests on different \gray\ spatial distribution hypotheses in both low (1$\sim$10 GeV) and high (>10 GeV) energy ranges.
The minimum value of $-\log(\cal L$) via the {\it gtlike} process corresponds to the largest likelihood value ${\cal L}_{\rm max}$ in the maximum likelihood method.
The test statistic is defined as TS = 2 log($\cal L_{\rm 1}/\cal L_{\rm 0}$), where $\cal L_{\rm0}$ is the likelihood value of the null hypothesis and $\cal L_{\rm1}$ is the likelihood of the hypothesis tested.
According to the theorems in \cite{Wilks1938AnnMathStatist..9..60,Protassov2002ApJ...571..545P}, the statistical significance $\sigma$ can be approximated from the distribution of the chi-square with n degrees of freedom, where n is the number of extra free parameters in the test hypothesis.
Following the definition and method in \cite{Lande2012}, the extension test statistic is quantified by $\rm TS_{ext}=2\log({\cal L}_{ext} /{\cal L}_{ps})$, where $\rm {\cal L}_{ext}$ is the maximum likelihood for the extended source model, and $\rm {\cal L}_{ps}$ for the point-like source model.
The \gray\ source is considered to be significantly extended only if $\rm TS_{ext}\ge$ 16.
And we apply the Akaike Information Criterion (AIC, \cite{Akaike1974ITAC...19..716A}), which is defined as AIC = $-2\log({\cal L}) + 2k$, where $k$ is the number of free parameters in the model.
The $\Delta$AIC we calculate is obtained by subtracting the AIC of the testing model from the AIC of the background model.
Thus the maximum value of $\Delta$AIC corresponds to the best-fit model like $\Delta$log$\cal L$ and TS values.

The residual \gray\ emissions below 10 GeV mostly overlap with the GMC W3 as seen in the upper panel of Fig.\ref{fig:resmap}.
We focus on the \gray\ emissions around GMC W3, firstly optimize the model of HB3 considering the two identified sources with spatial overlap.
We replace the model of HB3 (4FGL J0222.4+6156e) with a uniform disk with PowerLaw (PL) spectral type, and find a best-fit model by varying the center and radius of the disk (R.A. = 34.9\deg$\pm0.1$\deg, Dec. = 62.6\deg$\pm0.05$\deg, radius=0.7\deg$\pm0.05$\deg).
We add the optimized uniform disk into the background in the following analyses.
Then we replace the W3 (4FGL J0221.4+6241e) with several different models.
It is remarkable that the morphology of \gray\ emissions around W3 tends to be relatively symmetrical, so we prefer to test symmetrical models.
We test sources including the point-like source, radial Gaussian, elliptical Gaussian centered at the peak of the residual map ($p_{low}$, R.A. = 35.99\deg, Dec. = 62.01\deg) through changing the radii or long and short axes.
We also test two templates of CO due to the spatial coincidence with molecular gas (see Sect.\ref{sec:Gas}).
The models we test are all with the single power law (PL) spectral type.
As expected, the CO templates with asymmetric distribution with respect to $p_{low}$ is not the best-fit model, which is different from that of \cite{Katagiri2016}.
The best-fit model is the elliptical Gaussian ($\sigma$ to semimajor is 0.88$\pm0.12$, and the ratio of short and long axes, $r_{\rm b/a}$ is 0.4$\pm0.04$).

The lower panel in Fig.\ref{fig:resmap} gives an intriguing picture that the \gray\ emissions above 10 GeV concentrate in a small region where the YMCs and densest molecular gas are located.
Since we have optimized the model of HB3 in low energy band, we replace the model of HB3 (4FGL J0222.4+6156e) with the optimized uniform disk and find there is almost no change of the likelihood value.
So we also take the optimized uniform disk representing the model of HB3 as the background.
Then, we replace W3 (4FGL J0222.4+6156e) with a point-like source with a PL spectrum at the peak site ($p_{high}$, R.A. = 36.42$^\circ$, Dec. = 62.06$^\circ$) of residual map.
To test the extension of the point-like source, we replace it with CO templates and radial Gaussian centered at $p_{high}$.
Through changing the radius ($\sigma_{\rm disk}$) from 0.05\deg\ to 0.5\deg\ with a step of 0.05\deg, we find the best-fit radius is 0.15$^\circ$.
The best-fit radial Gaussian shows the maximum likelihood value.
However, the TS$_{\rm ext}$ value of 7 is less than 16 indicating that there is no significant extension.
We also use the elliptical Gaussian model fitted to the data in the 1-10 GeV energy range to replace W3, which shows very poor improvement with respect to the background.
This scenario is also a support for the energy-dependent distribution of gamma rays around W3 complex.

Thus, in the new model, we adopt the elliptical Gaussian (hereafter referred to as src A) and a point-like source near the cluster site (hereafter referred to as src B) to represent the \gray\ emissions around W3, and an optimized uniform disk for HB3.
We perform the binned likelihood analysis based on the new model in the energy range of 1 GeV to 300 GeV.
Compared to the model provided in the 4FGL-DR3 catalog, the improvement of TS of the new model approaches 85, corresponding to $\sim9\sigma$.
The TS values of the best-fit model and other templates are listed in Table.\ref{tab:likelihood}.
\begin{table*}
    \centering
    \caption{Results of spatial analyses for different models.}
    \begin{tabular}{cccccc}
        \hline\hline
        Energy Range & Model$^a$ & $\Delta$ log$\cal L$ & TS & D.o.f. & $\Delta$AIC \\
        \hline
        1-10 GeV & fermibkg$^b$ & 0 & 0 & 11 & 0 \\
        & 1ps$^c$ & 1317 & 2635 & 13 & 2631 \\
        & radial Gaussian & 1643 & 3286 & 14 & 3280 \\
        & CO(Dame) & 1694 & 3388 & 13 & 3384 \\
        & CO(MWISP) & 1671 & 3343 & 13 & 3339 \\
        & elliptical Gaussian & 1723 & 3446 & 15 & 3438 \\
        \hline
        10-300 GeV & fermibkg & 0 & 0 & 11 & 0 \\
        & 1ps & 14 & 28 & 13 & 24 \\
        & CO(Dame) & 13 & 26 & 13 & 22 \\
        & CO(MWISP) & 13 & 26 & 13 & 22 \\
        & radial Gaussian & 17 & 35 & 14 & 29 \\
        & elliptical Gaussian & 1 & 2 & 15 & -6 \\
        \hline
        1-300 GeV & fermibkg & 0 & 0 & 11 & 0 \\
        & fermi sources$^d$ & 1702 & 3404 & 15 & 3394 \\
        & elliptical Gaussian + 1ps & 1745 & 3489 & 17 & 3472 \\
        \hline
    \end{tabular}
    \label{tab:likelihood}
	\begin{tablenotes}
        \item $^a$ The spatial models here are for the source associated with W3.
	\item $^b$ No source associated with W3.
        \item $^c$ Single point source.
        \item $^d$ 4FGL J0222.4+6156e.
	\end{tablenotes}
\end{table*}

\subsection{Spectral analysis}
\label{sec:spectral_analysis}

To further study the influence of spectral type, we change the spectral type of src A to LogParabola (LogP), BrokenPowerLaw (BPL), and PLSupperExpCutoff (PLEC), respectively, and the spectra of the other two sources remain to be PL.
The formulae of these spectra are presented in Table~\ref{tab:spectral type}.
We find the best spectral type of src A through the binned likelihood analysis.
Then we keep the best choice for src A and remain the model of HB3 to be PL, and change the spectrum of src B to LogP, BPL, and PLEC, respectively, to find the best spectral function of src B.
We use a similar approach to find the best spectrum for the model of HB3.
\begin{table*}
    \centering
    \caption{Formulae for the \gray\ and particle spectral distributions.}
    \begin{tabular}{cccc}
         \hline
         & Name & Formulae & Free Parameters \\
         \hline
         \gray & PL & ${\rm d}N/{\rm d}E=N_0(E/E_0)^{-\Gamma}$ & $N_0, \rm \Gamma$ \\
         & LogP & ${\rm d}N/{\rm d}E=N_0(E/E_{\rm b})^{-\Gamma-\beta {\rm log}(E/E_{\rm b})}$ & $N_0,\Gamma,\beta$ \\
         & PLEC & ${\rm d}N/{\rm d}E=N_0(E/E_{\rm b})^{-\Gamma}{\rm exp}(-E/E_{\rm cut})$ & $N_0,\Gamma,E_{\rm cut}$\\
         & BPL & ${\rm d}N/{\rm d}E=$$\begin{array}{ll}
         \left\{
         \begin{aligned}
          N_0(E/E_{\rm b})^{-\Gamma_1}\ &:E<E_{\rm b}\\
          N_0(E/E_{\rm b})^{-\Gamma_2}\ &:E>E_{\rm b}\\
         \end{aligned}
         \right. \\
         \end{array}$ & $N_0,\Gamma_1,\Gamma_2,E_{\rm b}$\\
         \hline
         Particle & PL & $N(E)=A(E/E_0)^{-\alpha}$ & $A,\alpha$\\
         & ECPL & $N(E)=A(E/E_0)^{-\alpha}{\rm exp}(-(E/E_{\rm cut}))$ & $A,\alpha,\ \ E_{\rm cut}$\\
         & LogP & $N(E)=A(E/E_0)^{-\alpha-\beta {\rm log}(E/E_0)}$ & $A, \alpha,\ \beta$\\
         & BPL & $N(E)=$$\begin{array}{ll}
         \left\{
         \begin{aligned}
         A(E/E_{\rm 0})^{-\alpha_1}\ \ \ \ \ \ \ \ \ \ \ \ \ \ \ \ \ \ \ \ \ \ \ \ \ \ &:E<E_{\rm b}\\
         A(E_{\rm b}/E_0)^{\alpha_2-\alpha_1}(E/E_0)^{-\alpha_2}\ &:E>E_{\rm b}\\
         \end{aligned}
         \right. \\
         \end{array}$ & $A,\alpha_1,\ \alpha_2,\ E_{\rm b}$\\
         \hline
    \end{tabular}
    \label{tab:spectral type}
\end{table*}
The TS of each spectral type can be seen in Table~\ref{tab:best spectra}. 
Here we regard the PL spectrum as the null hypothesis.
The src A prefers a LogP spectrum and the other two sources will keep using the PL function in later analysis considering the small TS values.
\begin{table}
    \centering
    \caption{The likelihood-ratio test statistics (TS) for different spectral types favored over a PL null hypothesis.}
    \begin{tabular}{c|c|c|c|c}
    \hline
         Spectral Type & PL & LogP & BPL & PLEC \\
         \hline
         src A$^a$ & 0 & 27.4 & 16.6 & 26.0 \\
         src B$^b$ & 0 & 1.6 & 0.1 & 1.8 \\
         HB3$^c$ & 0 & 0.3 & 0.3 & 0.1 \\
         \hline
    \end{tabular}
    \label{tab:best spectra}
    \begin{tablenotes}
        \item $^a$ The spectral type of src B and HB3 are PL.
        \item $^b$ The spectral type of src A is LogP, the spectral type of HB3 is PL.
        \item $^c$ The spectral type of src A is LogP, the spectral type of src B is PL.
    \end{tablenotes}
\end{table}
For the GeV  \gray\ emission of src A, the photon indices are $\alpha=2.37\pm0.07$ and $\ \beta=0.25\pm0.04$.
The energy flux is $(1.94 \pm 0.05) \times 10^{-11}\rm \ erg\ cm^{-2}\ s^{-1}$,  corresponding to a \gray\ luminosity of $\sim 9.28 \times 10^{33}$ erg s$^{-1}$.
Here we adopt a distance of 2 kpc for W3 in the analysis.
For the \gray\ emission of HB3, we obtain the index of $\alpha=2.87\pm0.09$.
The energy flux is $(6.91 \pm 0.40) \times 10^{-12}\rm \ erg\ cm^{-2}\ s^{-1}$ corresponding to luminosity of $\sim 3.31 \times 10^{33}$ erg s$^{-1}$.
For the \gray\ emission from src B, the index is $2.19\pm0.25$, the flux is $\sim (4.06 \pm 2.27) \times 10^{-13}\rm \ erg\ cm^{-2}\ s^{-1}$ equivalent to a \gray\ luminosity of $\sim 1.94 \times 10^{32}$ erg s$^{-1}$.

\begin{figure*}
    \centering
    \includegraphics[width=0.45\linewidth]{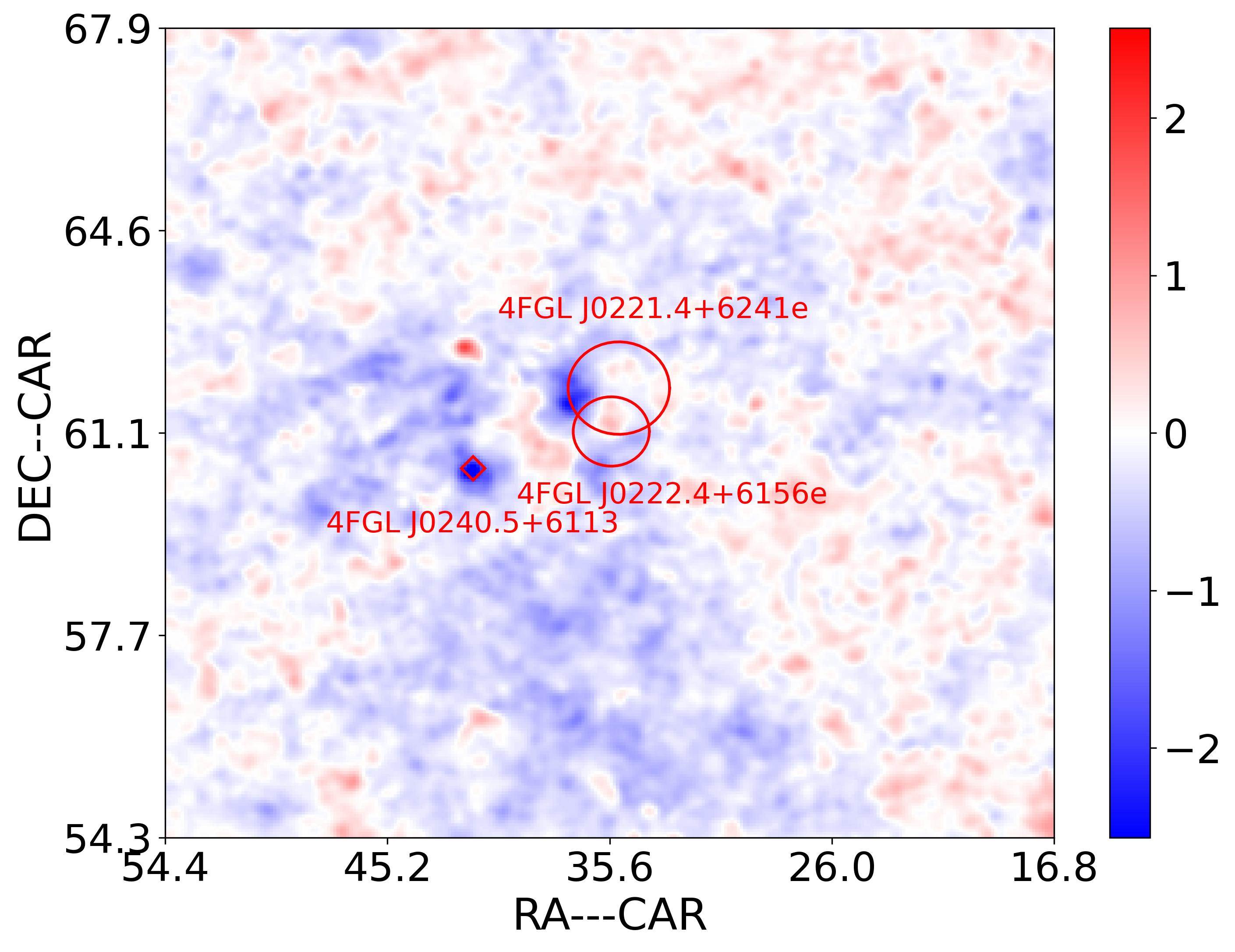}
    \includegraphics[width=0.45\linewidth]{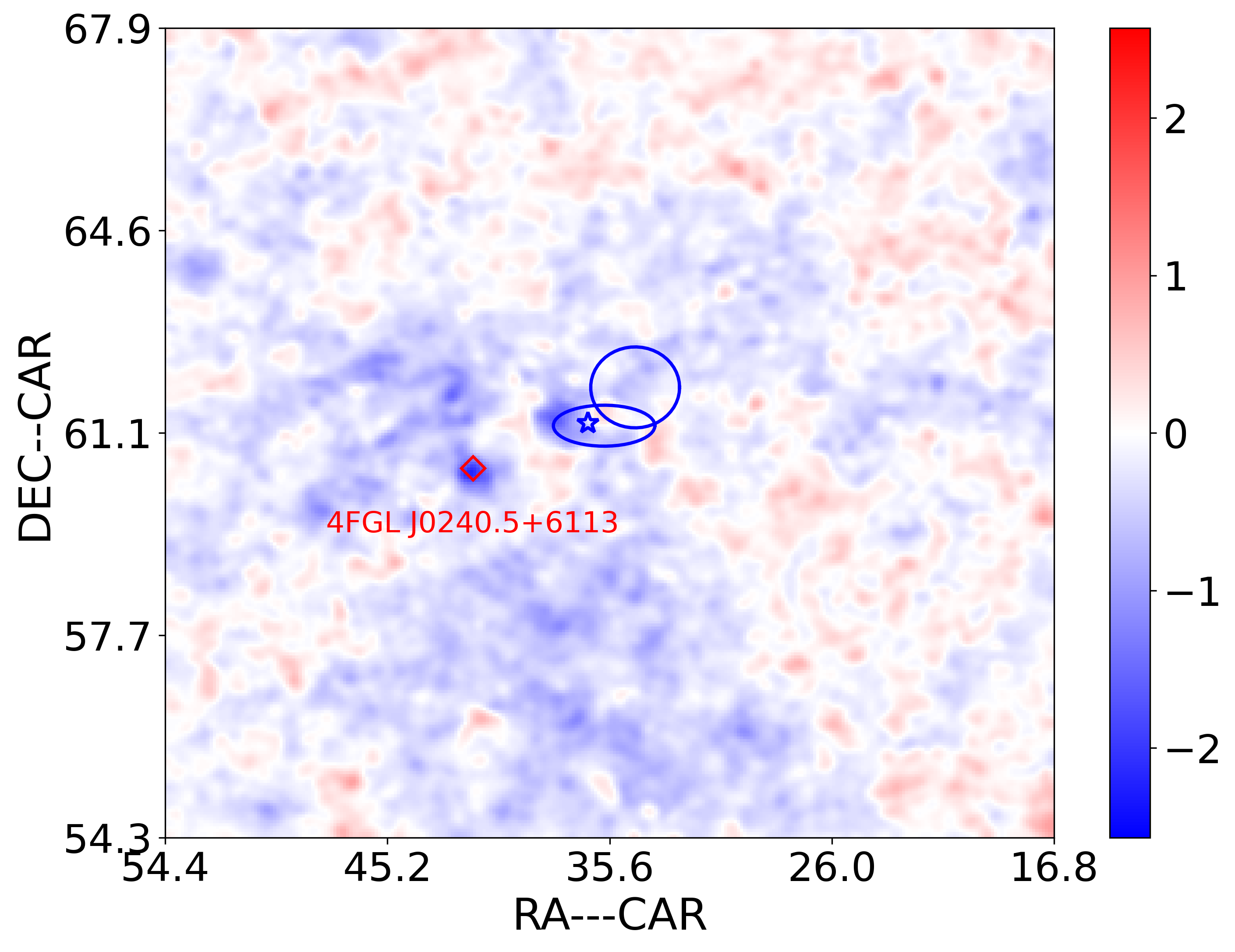}
    \caption{The PS maps for diagnostics of the goodness-of-fit.
    The left panel shows the PS map of the model only including sources in 4FGL catalog, and the right panel shows the PS map of the model adopted in this work.
    The red circles in the left panel represent two 4FGL sources, 4FGL J0221.4+6241e and 4FGL J0222.46156e, uesd to model the SNR HB3 and W3 complex by Fermi collaboration.
    The red diamond shows 4FGL J0240.5+6113, which is associated with X-ray binary LSI +61 303.
    And the three blue models in the right panel represent the three optimized sources used in this work.
    The limits of the colorbar are -2.57 and 2.57, respectively, corresponding to 3$\sigma$.}
    \label{fig:psmap}
\end{figure*}

Following the algorithm in \cite{Bruel2021A&A...656A..81B}, we use the \textit{gtpsmap.py}\footnote{\url{https://fermi.gscf.nasa.gov/ssc/data/analuysis/user}} to generate the PS map to test the goodness-of-fit of data-model agreement.
We generate the PS maps covering the ROI (14\deg\ $\times$ 14\deg) of the model that adopted in this work and the model which only includes 4FGL J0222.4+6156e for W3 and 4FGL J0221.4+6241e for HB3.
As we can see in the left panel in Fig.\ref{fig:psmap}, in the northeast of W3 there is an obvious positive residual where we add a point source into the background.
Also, an obvious deficit 2.3\deg\ away from W3 complex is coincident with 4FGL J0240.5+6113 which is associated with a fairly complex high mass X-ray binary system LSI +61 303.
We alter the spectral type from LogP adopted in 4FGL catalog to PL, BPL and PLEC, where BPL shows a better description of the model since the TS (sophisticated spectra relative to PL) is 24 higher than that of LogP.
Then we take the X-ray binary with a BPL spectrum as the background.
And for the large-scale clustering of residuals, with negative residuals mainly distributed along the Galactic plane and in a large blob centred at R.A. = 35\deg, Dec. = 57\deg, and positive residuals elsewhere, we replace the interstellar emission model (IEM) \textit{gll\_iem\_v07} of the standard version provided by Fermi collaboration with alternative IEM including dust template derived from \cite{Planck2011A&A...536A..13P} and IC emission template generated by GALPROP code with the GALDEF identification $^SY^Z6^R30^T150^C2$ used in \cite{Acero2016ApJS..223...26A} to check the impact of diffuse background.
The result also shows the scenario of clustering of residuals which cannot be eliminated by alternative IEM.
We use other three templates (CO, CO+\ion{H}{I}, and CO+\ion{H}{I}+\ion{H}{II}, in which CO is the same as that in \textit{gll\_iem\_07}, see details in Sec.\ref{sec:Gas}) to replace the dust template separately as well, of which the fitting results get even worse than that of dust template.
We therefore take the standard IEM as the diffuse emission model in this work.
Comparing the right to the left panel in Fig.\ref{fig:psmap}, the model in this work do reduce the deficit with respect to the model only including 4FGL catalog sources in the vicinity of W3 complex, which makes our results more plausible.
As for the deficit region around the X-ray binary a dedicated work is needed.

A pion-bump feature was found by \cite{Abdollahi2022ApJ...933..204A} for 4FGL J0222.4+6156e associated with W3.
We have replaced 4FGL J0222.4+6156e with two sources, src A and src B, of which src A contribute the most fluxes.
We refer to the method in \cite{Abdollahi2022ApJ...933..204A} to identify the pion-bump feature.
We use PSF3 and PSF2 type events in the energy band from 100 MeV to 1 GeV and \textit{edisp\_bins}=-3 to perform binned likelihood analysis which includes 10 logarithmically spaced bins.
The model obtained above is taken as the initial model to test the spectral curvature of src A.
We fit the initial model where the spectral type is PL at first, then replace the spectrum with LogP spectral type.
The improvement of the LogP model with respect to the PL one is performed by determining TS$_{\rm LogP}=2(\ln {\cal L}_{\rm LogP}-\ln {\cal L}_{\rm PL})$.
The resultant TS$_{\rm LogP}$ is 73 above 9 (which corresponds to 3$\sigma$ improvement for one additional degree of freedom), we then test a smoothly broken PL (SBPL),
\begin{equation}
\begin{aligned}
    dN/dE=N_0(E/E_0)^{-\Gamma_1}(1+(E/E_{\rm br})^{(\Gamma_2-\Gamma_1)/\alpha})^{-\alpha},
\end{aligned}
\end{equation}
where N$_0$ is the differential flux at $E_0$ = 300 MeV and $\alpha$ = 0.1.
This adds two additional degrees of freedom with respect to the PL model (the break energy $E_{\rm br}$ and a second spectral index $\Gamma_2$).
The improvement with respect to the PL one is determined by TS$_{\rm SBPL}$ = 2($\ln {\cal L}_{\rm SBPL}-\ln {\cal L}_{\rm PL}$).
We require TS$_{\rm SBPL}$ > 12 (implying a 3$\sigma$ improvement for two additional degrees of freedom) to keep the source in the significant energy break.
The tested TS$_{\rm SBPL}$ = 71.6 above 12 indicating a significant energy break which is the signature of pion-bump.
The break value of $430\pm40$ MeV is also compatible within 1$\sigma$ with 465$\pm 88$ MeV in \cite{Abdollahi2022ApJ...933..204A}.

\begin{figure}
    \centering
    \includegraphics[scale=0.45]{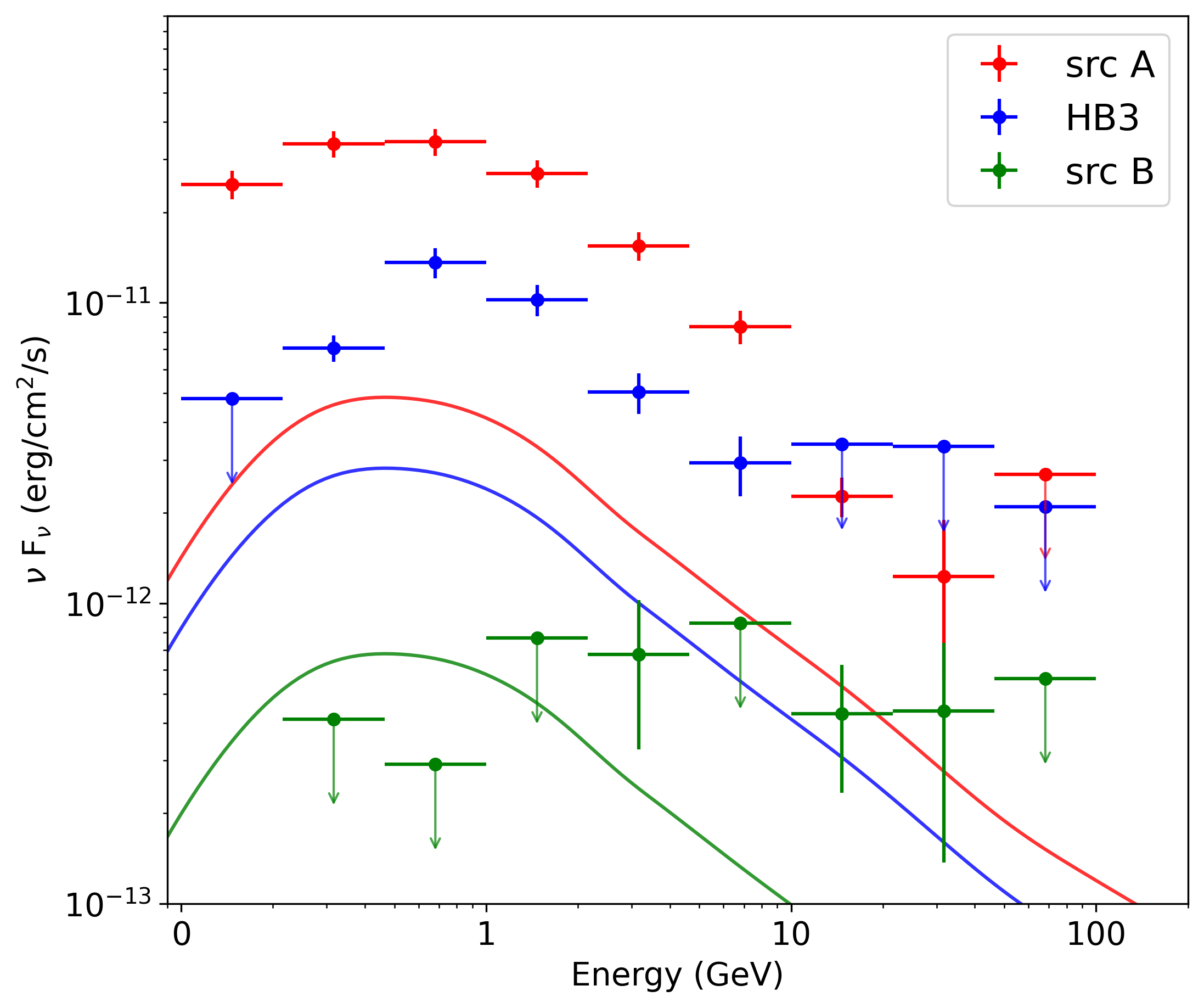}
    \caption{SEDs of the \gray\ emissions above 100 MeV of src A (red points), src B (green points) and HB3 (blue points).
    The solid lines represent the predicted \gray\ emissions of sources assuming the CR densities are the same as those measured locally by AMS-02 \citep{Aguilar2015}.
    Both the statistical and systematic errors are considered.
    }
    \label{fig:sed}
\end{figure}

We extract the spectral energy distributions (SEDs) of src A, src B and HB3 by performing the maximum likelihood fitting in logarithmically spaced energy bins above 100 MeV.
We calculate 2$\sigma$ upper limits for the energy bins in which the source's significance is lower than 2$\sigma$.
The extracted SEDs is shown in Figure \ref{fig:sed}.
The solid lines represent the predicted \gray\ emissions assuming the CR densities are equal to the ones at the Earth measured by AMS-02 \citep{Aguilar2015}.
The gas densities used to predict fluxes of \gray\ emissions are 230 cm$^{-3}$ and 27 cm$^{-3}$ for src A and HB3, respectively (See details in Sec.\ref{sec:Gas}).
For src B, we take gases where the density of H$_2$ is above 10$^{22}$ cm$^{-2}$ ($\sim 1.92 \times 10^4 \msun$) to estimate the predicted \gray\ flux from local CRs.
Although the extracted spectra of src A and HB3 do not get significantly harder with respect to the predicted ones, the fluxes have obvious excesses.
In order to estimate the systematic uncertainty associated with the imperfect modeling of the Galactic diffuse emission, we use eight alternative interstellar emission models (IEMs) generated through GALPROP\footnote{\url{http://glaprop.stanford.edu/}} (three variables: the CR source distributions of SNRs according to \cite{Case1998ApJ...504..761C} and of pulsars according to \cite{Lorimer2006MNRAS.372..777L}, the height of the CR propagation halo of 4 and 10 kpc, and the uniform spin temperature of 150 K and 10000 K) to repeat the analysis following \cite{Abdollahi2022ApJ...933..204A} and \cite{Acero2016ApJS..224....8A}.
In addition to the uncertainty of the IEM, we also consider the systematic errors due to the varied effective area of the detector.\footnote{\url{https://fermi.gsfc.nasa.gov/ssc/data/analysis/LAT_caveats.html}}
The total errors are obtained by adding the statistical and systematic errors in quadrature.

\section{Gas content around W3}
\label{sec:Gas}

CRs generate diffuse \gray\ emission by interacting with interstellar gas and radiation fields during their propagation through the Galaxy.
We can use the gas templates to derive the spatial and spectral information of the diffuse \gray\ emission if they have good spatial correlation. 
We study three different gas phases, i.e., the molecular hydrogen (H$_{2}$), the ionized hydrogen (\ion{H}{ii}), and the neutral atomic hydrogen (\ion{H}{i}), in the vicinity of the W3 region.

\begin{figure*}
    \centering
    \includegraphics[scale=0.24]{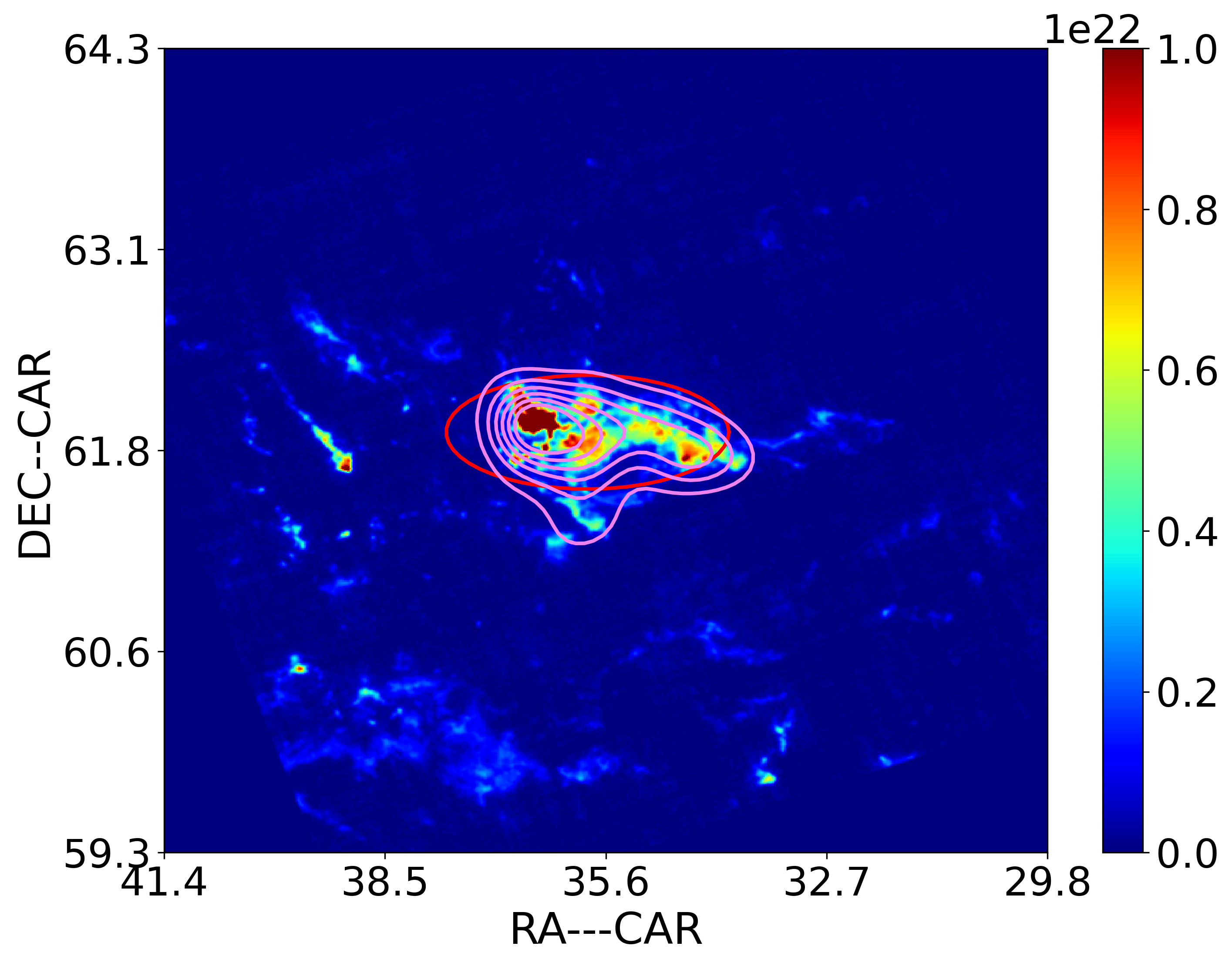}
    \includegraphics[scale=0.24]{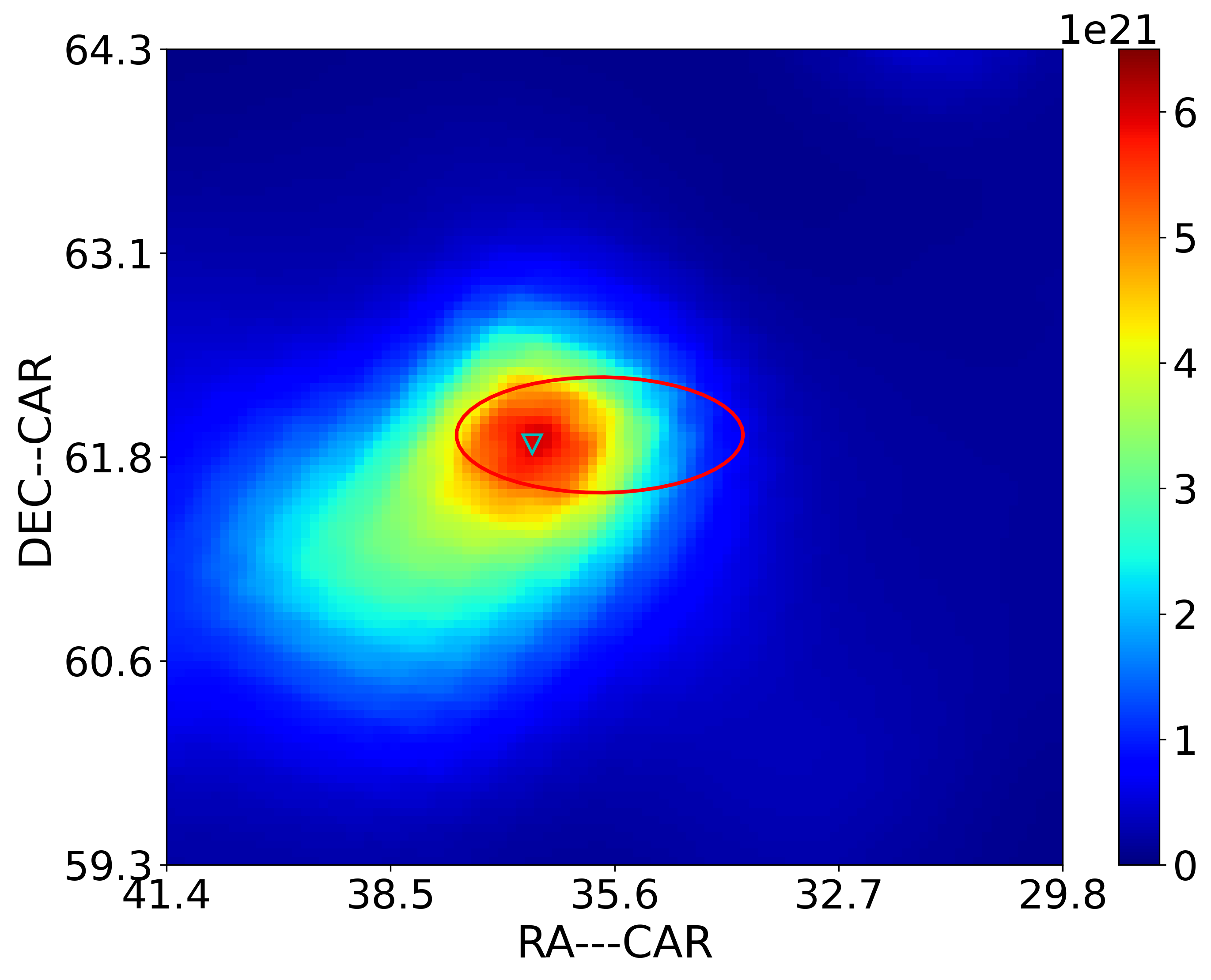}
    \includegraphics[scale=0.24]{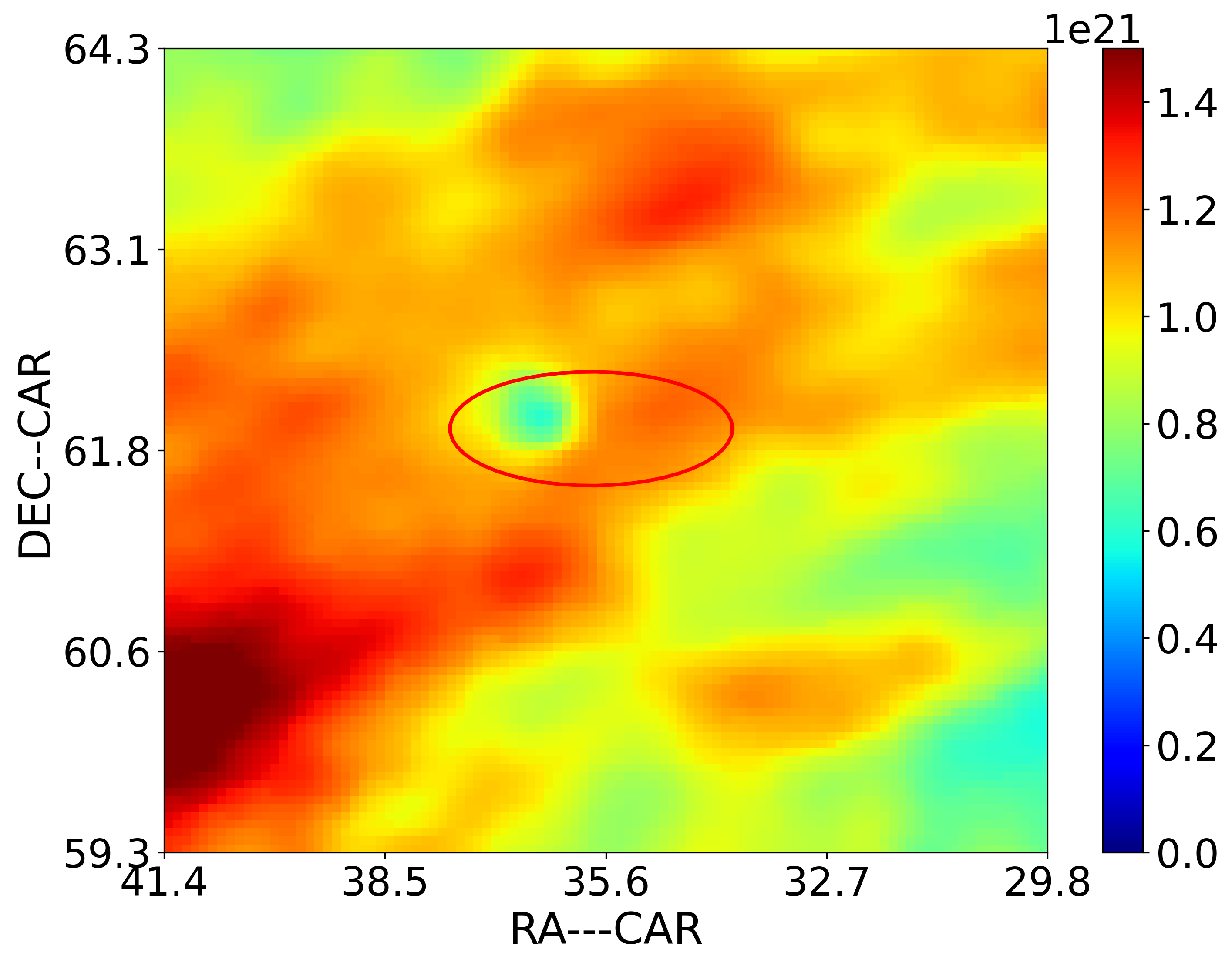}
    \caption{Maps of gas column densities for three gas phases. The left panel shows the H$_{2}$ column density derived from the CO MWISP data, where the violet contours represent the distributions of CO Dame. The middle panel shows the \ion{H}{ii} column density derived from the \planck\ free-free map assuming the effective density of electrons $n_{\rm e}=2~\rm cm^{-3}$. The right panel shows the map of \ion{H}{i} column density derived from 21-cm all-sky survey. The red ellipse indicates the src A, and the inverted triangle represents the $\ion{H}{II}$ region IC 1795.}
    \label{fig:gas}
\end{figure*}

We adopt the observations of $^{12}$CO line emissions toward W3 with the Purple Mountain Observatory Delingha 13.7 m millimeter-wavelength telescope \citep{Zuo2011}, which is a part of the Milky Way Image Scroll Painting (MWISP) survey project \footnote{\url{http://www.radioast.nsdc.cn/yhhjindex.php}} to trace the molecular hydrogen $\rm H_2$ (referred to as CO MWISP).
We also generate the composite CO survey data from \cite{Dame2001} (referred to as CO Dame) which is the same as that in \cite{Katagiri2016}.
We analyze the $\rm H_2$ by the standard assumption of a linear relationship between the column density of molecular hydrogen, $N_{\rm H_{2}}$, and the velocity-integrated brightness temperature of 2.6 mm line of the carbon monoxide (CO), W$_{\rm CO}$, i.e. $N({\rm H_{2}}) = X_{\rm CO} \times W_{\rm CO}$ \citep{Lebrun1983,Dame2001}.
$X_{\rm CO}$ is the empirical conversion factor which is set to be $\rm 2.0 \times 10^{20}\ cm^{-2}\ K^{-1}\ km^{-1}\ s$ according to \cite{Dame2001,Bolatto2013}.
$\rm H_2$ distribution shown in the left panel of Fig.\ref{fig:gas} is derived according to the range of radial velocity, -44.2 $\sim$ -33.8 $\rm km\ s^{-1}$ \citep{Katagiri2016}.

A large diffuse $\ion{H}{II}$ region IC 1795 is located between W3 Main and W3(OH).
The galactic coordinate (R.A. = 133.86\deg, Dec. = 1.15\deg) of IC 1795 implies it is toward the outer galaxy where relatively less gas is existing.
And other HII regions are not discovered in the direction.
So we consider the same distance of HII as that of W3 complex.
To trace $\ion{H}{II}$, we utilize the free-free emission map which is derived from Planck, WMAP and 408 MHz radio observations \citep{Planck2016}.
We firstly use the conversion factor in Table 1 of \cite{Finkbeiner2003} to convert the emission measure (EM) into free-free intensity ($I_{\nu}$). Then we calculate the $\ion{H}{II}$ column density by using the Eq.(5) of \cite{Sodroski1997},
\begin{equation}
 \begin{aligned}
N_{\ion{H}{II}} = &1.2 \times 10^{15}\ {\rm cm^{-2}} \left(\frac{T_{\rm e}}{1\ \rm K}\right)^{0.35} \left(\frac{\nu}{1\ \rm GHz}\right)^{0.1}\left(\frac{n_{\rm e}}{1\ \rm cm^{-3}}\right)^{-1} \\
&\times \frac{I_{\nu}}{1\ \rm Jy\ sr^{-1}},
 \end{aligned}
\end{equation}
where the conversion frequency $\nu$ is at 353 GHz, and the electron temperature $T_{\rm e} $ = 8000 K.
We choose 2 cm$^{-3}$ recommended by \cite{Sodroski1997} to derive the $\ion{H}{II}$ column density for regions outside the solar circle.
The derived gas column density map is shown in the middle panel of Fig.\ref{fig:gas}.

The $\ion{H}{I}$ data is taken from $\ion{H}{I}$ 4$\pi$ survey (HI4PI), a data-cube of 21-cm all-sky Galactic $\ion{H}{I}$ observations \citep{HI4PI2016}.
We derive the $\ion{H}{I}$ column density $N_{\ion {H}{I}}$ via the expression
\begin{equation}
N_{\ion{H}{I}} = -1.83 \times 10^{18}\ {\rm cm^{-2}}\ T_{\rm s}\int {\rm d}\upsilon\ {\rm ln} \left(1-\frac{T_{\rm B}}{T_{\rm s}-T_{\rm bg}}\right),
\end{equation}
where $T_{\rm B}$ and $T_{\rm bg} \approx 2.66$ K are the brightness temperature of the $\ion{H}{I}$ and the brightness temperature of the cosmic microwave background radiation at 21 cm, respectively.
In case of $T_{\rm B} > T_{\rm s} - 5\ \rm K$, we truncate $T_{\rm B}$ to $T_{\rm s} - 5\ \rm K$ following \cite{Ackermann2012ApJ...750....3A}, in which a uniform spin temperature $T_{\rm s}$ is 150 K.
The integral velocity range is the same as that of CO gas.
The column density map of \ion{H}{I} is shown in the right panel of Fig.\ref{fig:gas}, where the red ellipse indicates the src A and the inverted triangle represents the $\ion{H}{II}$ region IC 1795.

The total mass is calculated via the expression
\begin{equation}
    M_{\rm H} = m_{\rm H} N_{\rm H} A_{\rm angular} d^{2}
\end{equation}
in which $m_{\rm H}$ is the mass of hydrogen atom, $N_{\rm H}=N_{\rm {\ion{H}{I}}}+2N_{\rm H_2}+N_{\ion{H}{II}}$ is the column density of atoms in each pixel, $A_{\rm angular}$ is the angular area and $d$ is the distance.
We take the model of src A as an ellipsoid and HB3 as a sphere.
Then we get the volume of ellipsoid by $V = \frac{4}{3}\pi r_1r_2^2$, where $r_1$ is the semimajor and $r_2$ is the semiminor.
$r_1$, $r_2$ are calculated according to $r = d\times \theta$(rad) where $d$ and $\theta$ are the distance to Earth and the observational scale of projection on the sky, respectively.
When calculating the mass of H$_2$, we use the data from CO MWISP owing to the better spatial and velocity resolution with respect to CO Dame.
Since relative small mass of $\ion{H}{I}$, less than ten percent of the sum of H$_2$ and $\ion{H}{II}$, whether or not HI is considered has little effect on the fitting proton spectrum.
Coupled with the spatial coincidence with molecular and ionized hydrogen, we adopt H$_2$ and $\ion{H}{II}$ to derive the gas number density of src A.
The derived gas number density of src A is $\sim 230\ \rm cm^{-3}$.
For HB3, because of the comparable mass of H$_2$, $\ion{H}{II}$ and $\ion{H}{I}$, we account for the three phase gases to derive the number density, $\sim$27 cm$^{-3}$.
The masses of gases and derived number densities are shown in Tab.~\ref{tab:mass}.
\begin{table}
    \centering
    \caption{The masses and number densities of gases corresponding to the regions of src A and HB3.}
    \begin{tabular}{c|c|c|c}
        \hline
        gas phases & region & mass $(10^4\msun)$ & number density (cm$^{-3}$) \\
        \hline
        H$_2$+\ion{H}{II} & src A & 13.77 & 230 \\
        H$_2$+\ion{H}{I}+\ion{H}{II} & HB3 & 8.07 & 27 \\
        \hline
    \end{tabular}
    \label{tab:mass}
\end{table}

\section{The origin of gamma-ray emission}
\label{sec:origin}

To clarify the possible radiation mechanisms of the \gray\ emissions around W3, we fit the SEDs using the $\it Naima$ package\footnote{\url{https://naima.readthedocs.io/en/latest/index.html}} \citep{Zabalza2015}.
$\it Naima$ allows for a Markov Chain Monte Carlo (MCMC) fitting through the use of $\it emcee$ \citep{Foreman2013}.

\subsection{src A}

\begin{figure*}
    \centering
    \includegraphics[scale=0.42]{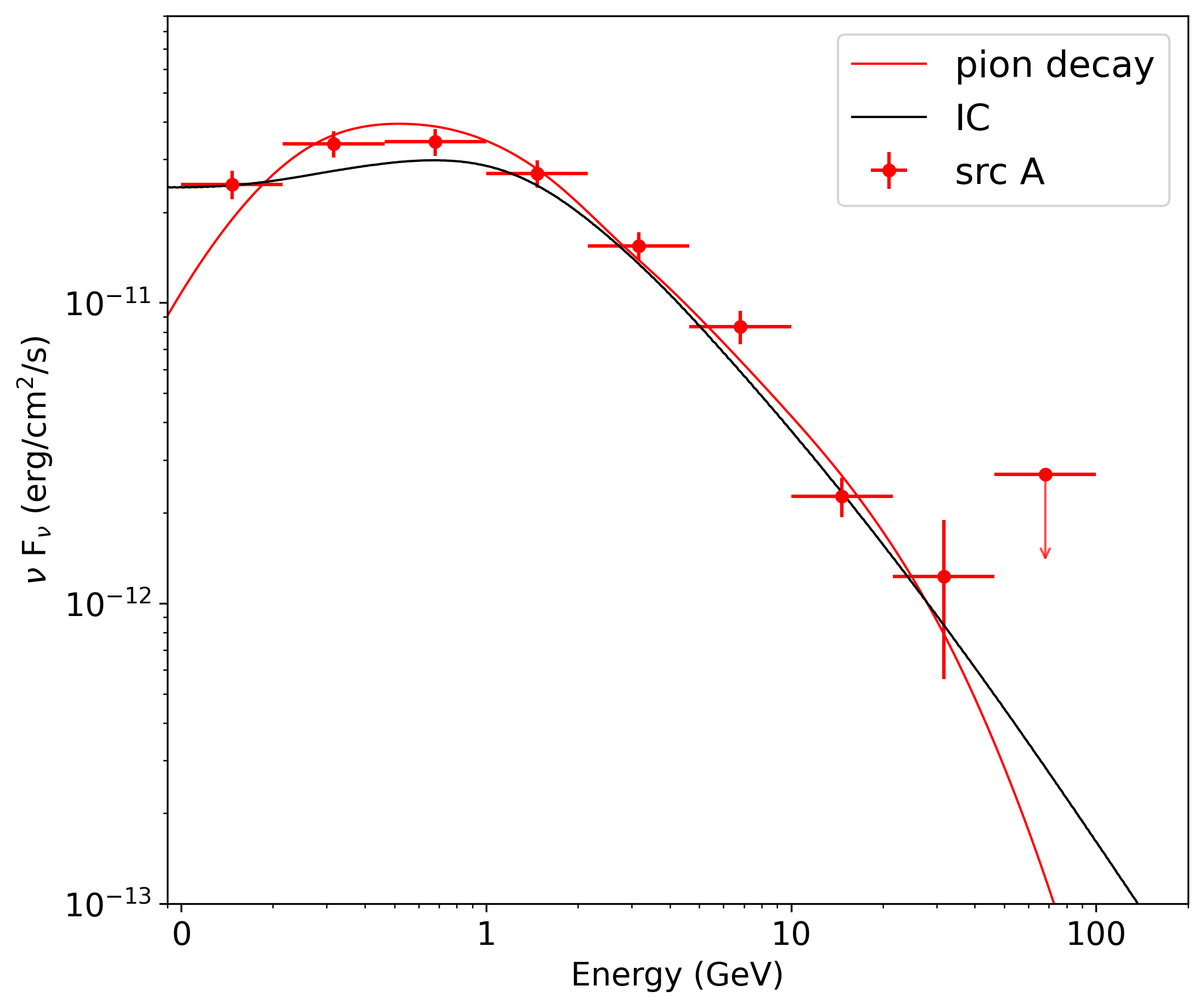}
    \includegraphics[scale=0.42]{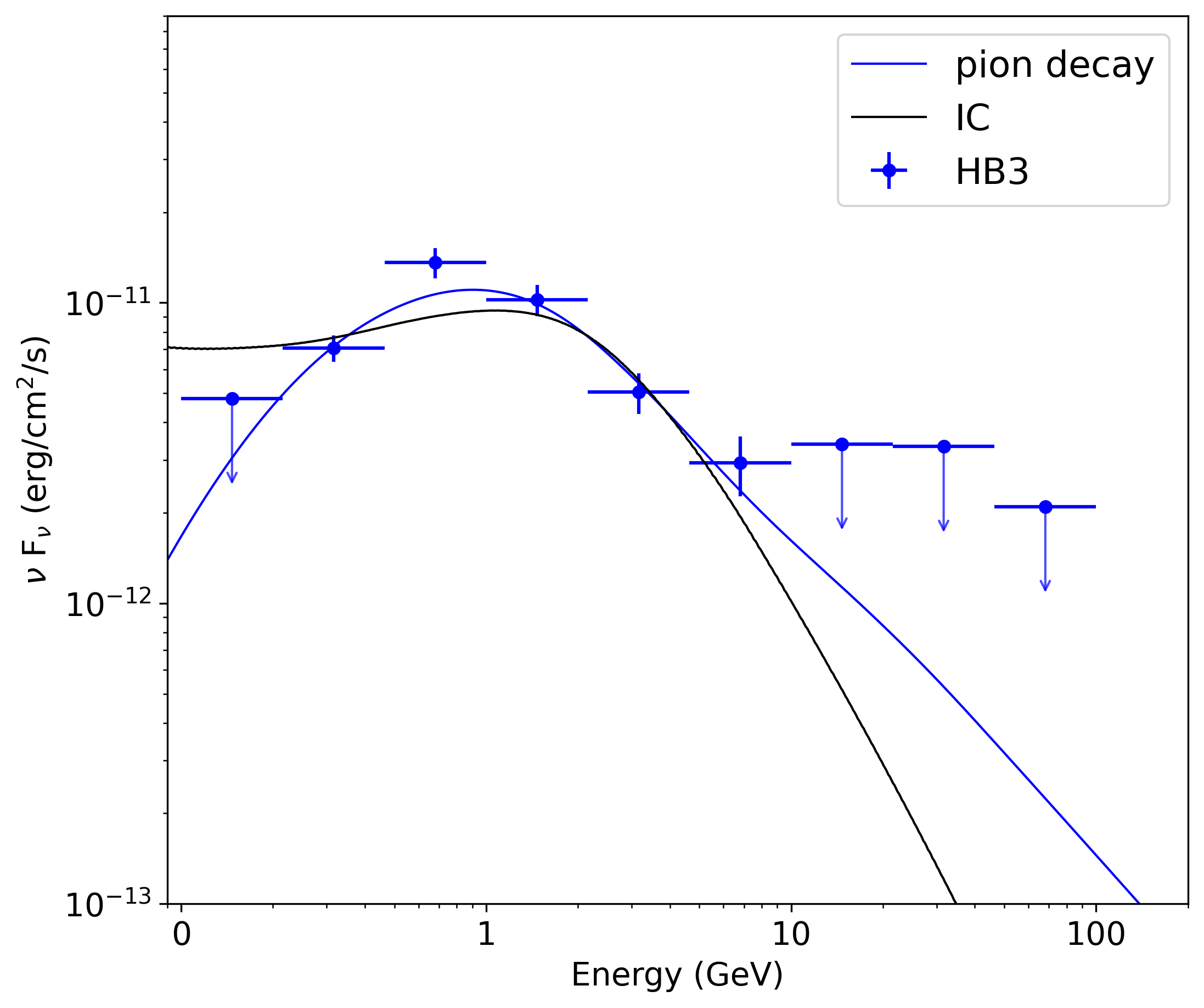}
    \caption{The SEDs of src A and HB3 are the same as that in Fig.\ref{fig:sed}. The colored and black solid lines are the fitting lines for the pion decay model and the IC model, respectively.
    }
    \label{fig:fit_W3}
\end{figure*}

We note that there is a clear spatial correlation between the extended GeV \gray\ emission of src A and the molecular hydrogen gas, though the densest region of the molecular gas deviates from the peak of the \gray\ emissions and the model centroid.
The pion-bump feature (See details in Sec \ref{sec:spectral_analysis}) indicates a possible hadronic origin for the \gray\ emissions.
We firstly assume the \gray\ emissions are produced from proton-proton inelastic interactions (PP) of the CRs with ambient gas via the pion-decay process.
We assume the parent protons have different spectral distributions including PowerLaw (PL), LogParabola (LogP), Exponential Cutoff PowerLaw (ECPL) and Broken PowerLaw (BPL), and use the \gray\ production cross-section of \cite{Kafexhiu2014} of the proton-proton collision (PP) model to fit the \gray\ data points.
The gas number density of src A is set to 230 cm$^{-3}$.
We adopt the ECPL spectrum which shows the maximum likelihood value among those spectral models of parent protons.
As shown in the left panel of Fig.\ref{fig:fit_W3}, the red data points are the SED of src A which is the same as that in Fig.~\ref{fig:sed}, and the red solid line is the fitting result of the PP model.
The derived indices are $\alpha = 2.66 \pm 0.11, E_{\rm cut} = 80^{+30}_{-20}$ GeV, and the total energy of protons is $W_{\rm p} = (1.89^{+0.10}_{-0.08}\times 10^{48}$ erg above 2 GeV.

We also consider the leptonic origin of the \gray\ emissions with the inverse Compton (IC) scattering scenario.
The seed photon fields of IC scattering include Cosmic Microwave Background (CMB), infrared radiation (IR), and optical radiation based on the model by \cite{Popescu2017MNRAS.470.2539P}.
We use the formalism in \cite{Khangulyan2014} with different electron distributions to fit the SED.
The black solid line in Figure~\ref{fig:fit_W3} is the fitting result of the IC model.
The best-fit electron spectrum is BPL, the derived indices are $\alpha_1 = 0.80 \pm 0.20, \alpha_2 = 4.47^{+0.20}_{-0.14}$, and the total energy is $W_{\rm e} = (1.63 \pm 0.16)\times 10^{50}$ erg for the electrons above 2 GeV, which is slightly more than 10 percent of typical energy of a supernova explosion.
From the perspective of energy budget, the fitting results seem to support the hadronic origin, but we cannot formally rule out the leptonic scenario as a result of potential enhanced radiation fields provided by stars in W3 complex.

\subsection{HB3}

We fit the SED of HB3 with the PP and IC model with various particle spectra.
The BPL distribution shows the maximum likelihood value for both PP and IC model.
For PP model, the proton index $\alpha_1=1.1^{+0.4}_{-0.5}, \alpha_2=3.2^{+0.7}_{-0.3}, E_{\rm b}=18\pm3{\rm GeV}, W_{\rm p}=4.0\pm0.3\times10^{48}$erg for protons above 2 GeV.
For IC model, the derived index $\alpha_1=0.7\pm0.3,\alpha_2=5.2^{+1.0}_{-0.6}, E_{\rm b}=12\pm1.3 {\rm GeV}, W_{\rm e}=4.2\pm0.3\times10^{49}$erg for electrons above 2 GeV.
The blue data points in the right panel of Fig.\ref{fig:fit_W3} are the SED of HB3 which is the same as that in Fig.\ref{fig:sed}, the blue and black solid line are the fitting results of PP and IC models, respectively.
\cite{Boumis2022} obtained the energy of HB3 SNR about $(3.70 \pm 1.50$) $\times$ 10$^{51}$ erg assuming that the SNR is in the Pressure Driven Snowplow (PDS) evolution phase.
\cite{Lazendic2006} estimated the explosion energy of $(3.40 \pm 1.50) \times 10^{50}$ erg through X-ray emission assuming the SNR is in the adiabatic stage.
Regardless of whether CRs being hadrons or leptons, the explosion of the supernova is able to offer sufficient energy.

\subsection{src B}

Src B is a new point source that is not included in the 4FGL catalog at the position of the cluster W3 Main.
Due to only few data points, the parent particle spectral parameters is poorly limited.
However, the flux of src B indeed exists excess when the photon energy is above 10 GeV as can be seen from Fig.\ref{fig:sed}, which conforms to the morphology analysis.
And the data points of the SED shows a slight hardening trend relative to the predicted \gray\ emissions.
This scenario makes the model in which we resolve W3 into two components in this work more plausible.

\section{Discussion and conclusion}
\label{sec:conclusion}

We present a detailed analysis based on about fourteen years of \fermi\ data toward the W3 complex.
Through the spatial analysis, we resolve the excess of \gray\ emissions from W3 into two components, an elliptical Gaussian
covering the molecular gas and a point-like source near the cluster W3 Main.

For src A, the pion-bump feature and better fitting result of pion-decay model, and the spatial consistence of the \gray\ emission with the molecular gas naturally favor the hadronic origin.
There are two promising celestial objects of CRs around the W3, the SNR HB3 located at the northwest of W3 and YMCs lying in the high-density layer.
Assuming a single massive star powering kinetic energy of $\sim$3 $\times\ 10^{48}$ erg within $\sim$ 1 Myr \citep{Ezoe2006}, the whole clusters harbouring $\sim$ 100 OB stars \citep{Yamada2021} can produce a total energy of $6\times10^{50}$erg taking into account the age of clusters of 2 Myr \citep{Kiminki2015ApJ...813...42K}.
It is significantly higher than the derived energy of parent protons $(1.89^{+0.10}_{-0.08})\times 10^{48}$ erg under the hadronic scenario.
So these YMCs have the capability to power the CRs.
And no pulsars and pulsar winds nebulae are detected nearby the GMC.
If we only consider YMCs as the origin of \gray\ emission from src A, the derived energy of CR protons is $\sim 10^{48}$ erg and the wind power of the whole clusters is about 10$^{37}$ erg s$^{-1}$.
So the confinement time of CRs can reach 10$^{12}$ s assuming a acceleration efficiency of 10\%.
Taking into account of the distance of the YMC W3 Main to the rim of \gray\ emissions about 37 pc (1.08$^\circ$ in 2 kpc).
The diffuse coefficient can be estimated by $D = \frac{l^2}{4T}$ as $\sim 3.23\times 10^{27}$ cm$^2$ s$^{-1}$.
Actually, due to the approximately symmetric morphology and the large extension of the \gray\ emission of W3, it is possibly to be that the CRs are accelerated near the peak of the \gray\ emission and propagate around.
Thus the interactions between the winds of clusters and shock produced by SNR could be the natural explanation of the origin of parent CRs.
If so, the diffuse coefficient should be more than 1.29 $\times 10^{27}{\rm cm^2s^{-1}}$ considering that the diffuse time is smaller than the age of the SNR HB3 ($\sim$3 $\times 10^{4}$ yr).
The two assumptions both obtain a slow diffuse coefficient $\sim$2 magnitudes smaller than that in the galactic plane \citep{Strong1998ApJ...509..212S}, suggesting that there may exist strong magnetic fields or turbulence probably due to the gas density gradient \citep{Crutcher2012ARA&A..50...29C}.
Nevertheless, accounting for the index of proton spectrum from src A of 2.6 is close to the local CR spectral index, we can not rule out that these emissions come from the large-scale background CRs population interacting with extra gas.

The new point-like source src B lying at the cluster W3 Main is only bright above several GeV, which could be caused by the obstruction of CR penetration due to the very high density of the gas.
It may be similar to that in \cite{Yang2023NatAs...7..351Y}.
The difference in photon indices between src A and src B also implies that \gray\ emissions of two sources have different origins (If the spectrum of src A is a PL, the index will be 2.87$\pm$0.03.).
The spatial coincidence between the \gray\ emission and W3 Main, and similar photon index with other YMC systems \citep{Ge2022,Liu2022} indicate it may be directly powered by the star cluster.
The W3 Main cluster has enough luminosity $\sim$$1.90\times10^{36}$ erg s$^{-1}$ (considering one fifth of total wind power for W3 Main) to drive the source $\sim$$1.94\times10^{32}$ erg s$^{-1}$.
It may be another YMC system that powers high-energy CRs interacting with surrounding gas to produce \gray\ emissions.

And one can suppose that the \gray\ emissions around W3, src A and src B, are both produced by the proton-proton interaction between relativistic hadrons accelerated by shock from SNR and ambient gas similar to that of \cite{Katagiri2016}.
The higher-energy \gray\ emission in Src B is further than the lower-energy \gray\ emission in src A with respect to the SNR, which may be due to the the energy-dependent propagation where high-energy CRs can propagate further.
While, in that case, high-energy CRs should collide with almost the entire molecular gas and illuminate the whole molecular gas, which is contradictory with the observed higher-energy \gray\ emissions shown in Fig~\ref{fig:resmap} lower panel.
The possible explanation is that the \gray\ emissions produced by high-energy CRs is relatively weak so it was not found in the TS map or counts map, but only the area with highest gas density is bright above 10 GeV.
It should be noted that although MWISP project tracing CO uses more advanced instruments, the CO data from MWISP project we used is actually different from the CO in background \textit{gll\_iem\_v07} from \cite{Dame2001}, between which the inconsistency may have an impact on our conclusion.
In addition, the clustering of residuals in Fig.\ref{fig:psmap} may have an effect upon our result, which need more subtle and detailed gas analysis.
Further deep investigations of energy-dependent morphology are needed with great caution.

\section*{Acknowledgements}

This work is supported by the National Natural Science Foundation of China (Grant No.12133003, 12103011, and U1731239), Guangxi Science Foundation (grant No. AD21220075 and 2024GXNSFBA010375).
This work is also supported by the Guangxi Talent Program ("Highland of Innovation Talents").
Rui-zhi Yang is supported by the National Natural
Science Foundation of China under grants 11421303, 12041305, and the national youth thousand talents program in China.
This research made use of the data from the Milky Way Imaging Scroll Painting (MWISP) project, which is a multi-line survey in $^{12}$CO/$^{13}$CO/C$^{18}$O along the northern galactic plane with PMO-13.7m telescope.
MWISP was sponsored by National Key R\&D Program of China with grant 2017YFA0402701 and by CAS Key Research Program of Frontier Sciences with grant QYZDJ-SSW-SLH047.

\section*{Data availability}
The \fermi\ data used in this work is publicly available, which is provided online by the NASA-GSFC Fermi Science Support Center\footnote{\url{ https://fermi.gsfc.nasa.gov/ssc/data/access/lat/}}.
A MWISP open data proposal is required to access the CO line data used in this work, and the form could be downloaded from \url{http://english.dlh.pmo.cas.cn/op/odp/}.
We also make use of the CO data\footnote{\url{ https://lambda.gsfc.nasa.gov/product/}}.
The data from \planck\ legacy archive\footnote{\url{ http://pla.esac.esa.int/pla/\#home}} are used to derive the \ion{H}{II}.
The \ion{H}{i} data are from the HI4PI\footnote{\url{http://cdsarc.u-strasbg.fr/viz-bin/qcat?J/A+A/594/A116}}.

\bibliographystyle{mnras}
%\bibstyle{aa}
%\bibliographystyle{plain}
\bibliography{ref}
\bibliography{ms}

\begin{thebibliography}{}
\makeatletter
\relax
\def\mn@urlcharsother{\let\do\@makeother \do\$\do\&\do\#\do\^\do\_\do\%\do\~}
\def\mn@doi{\begingroup\mn@urlcharsother \@ifnextchar [ {\mn@doi@}
  {\mn@doi@[]}}
\def\mn@doi@[#1]#2{\def\@tempa{#1}\ifx\@tempa\@empty \href
  {http://dx.doi.org/#2} {doi:#2}\else \href {http://dx.doi.org/#2} {#1}\fi
  \endgroup}
\def\mn@eprint#1#2{\mn@eprint@#1:#2::\@nil}
\def\mn@eprint@arXiv#1{\href {http://arxiv.org/abs/#1} {{\tt arXiv:#1}}}
\def\mn@eprint@dblp#1{\href {http://dblp.uni-trier.de/rec/bibtex/#1.xml}
  {dblp:#1}}
\def\mn@eprint@#1:#2:#3:#4\@nil{\def\@tempa {#1}\def\@tempb {#2}\def\@tempc
  {#3}\ifx \@tempc \@empty \let \@tempc \@tempb \let \@tempb \@tempa \fi \ifx
  \@tempb \@empty \def\@tempb {arXiv}\fi \@ifundefined
  {mn@eprint@\@tempb}{\@tempb:\@tempc}{\expandafter \expandafter \csname
  mn@eprint@\@tempb\endcsname \expandafter{\@tempc}}}

\bibitem[\protect\citeauthoryear{{Abdo} et~al.,}{{Abdo}
  et~al.}{2009}]{Abdo2009}
{Abdo} A.~A.,  et~al., 2009, \mn@doi [Astroparticle Physics]
  {10.1016/j.astropartphys.2009.08.002}, \href
  {https://ui.adsabs.harvard.edu/abs/2009APh....32..193A} {32, 193}

\bibitem[\protect\citeauthoryear{{Abdollahi} et~al.,}{{Abdollahi}
  et~al.}{2020}]{4FGL}
{Abdollahi} S.,  et~al., 2020, \mn@doi [\apjs] {10.3847/1538-4365/ab6bcb},
  \href {https://ui.adsabs.harvard.edu/abs/2020ApJS..247...33A} {247, 33}

\bibitem[\protect\citeauthoryear{{Abdollahi} et~al.,}{{Abdollahi}
  et~al.}{2022a}]{4FGL-DR3}
{Abdollahi} S.,  et~al., 2022a, \mn@doi [\apjs] {10.3847/1538-4365/ac6751},
  \href {https://ui.adsabs.harvard.edu/abs/2022ApJS..260...53A} {260, 53}

\bibitem[\protect\citeauthoryear{{Abdollahi} et~al.,}{{Abdollahi}
  et~al.}{2022b}]{Abdollahi2022ApJ...933..204A}
{Abdollahi} S.,  et~al., 2022b, \mn@doi [\apj] {10.3847/1538-4357/ac704f},
  \href {https://ui.adsabs.harvard.edu/abs/2022ApJ...933..204A} {933, 204}

\bibitem[\protect\citeauthoryear{{Acero} et~al.,}{{Acero}
  et~al.}{2016a}]{Acero2016ApJS..223...26A}
{Acero} F.,  et~al., 2016a, \mn@doi [\apjs] {10.3847/0067-0049/223/2/26}, \href
  {https://ui.adsabs.harvard.edu/abs/2016ApJS..223...26A} {223, 26}

\bibitem[\protect\citeauthoryear{{Acero} et~al.,}{{Acero}
  et~al.}{2016b}]{Acero2016ApJS..224....8A}
{Acero} F.,  et~al., 2016b, \mn@doi [\apjs] {10.3847/0067-0049/224/1/8}, \href
  {https://ui.adsabs.harvard.edu/abs/2016ApJS..224....8A} {224, 8}

\bibitem[\protect\citeauthoryear{{Ackermann} et~al.,}{{Ackermann}
  et~al.}{2012a}]{Ackermann2012ApJS..203....4A}
{Ackermann} M.,  et~al., 2012a, \mn@doi [\apjs] {10.1088/0067-0049/203/1/4},
  \href {https://ui.adsabs.harvard.edu/abs/2012ApJS..203....4A} {203, 4}

\bibitem[\protect\citeauthoryear{{Ackermann} et~al.,}{{Ackermann}
  et~al.}{2012b}]{Ackermann2012ApJ...750....3A}
{Ackermann} M.,  et~al., 2012b, \mn@doi [\apj] {10.1088/0004-637X/750/1/3},
  \href {https://ui.adsabs.harvard.edu/abs/2012ApJ...750....3A} {750, 3}

\bibitem[\protect\citeauthoryear{Aguilar et~al.,}{Aguilar
  et~al.}{2015}]{Aguilar2015}
Aguilar M.,  et~al., 2015, \mn@doi [Phys. Rev. Lett.]
  {10.1103/PhysRevLett.114.171103}, 114, 171103

\bibitem[\protect\citeauthoryear{{Aharonian}, {Yang}  \& {de O{\~n}a
  Wilhelmi}}{{Aharonian} et~al.}{2019}]{Aharonian2019NatAs...3..561A}
{Aharonian} F.,  {Yang} R.,   {de O{\~n}a Wilhelmi} E.,  2019, \mn@doi [Nature
  Astronomy] {10.1038/s41550-019-0724-0}, \href
  {https://ui.adsabs.harvard.edu/abs/2019NatAs...3..561A} {3, 561}

\bibitem[\protect\citeauthoryear{{Akaike}}{{Akaike}}{1974}]{Akaike1974ITAC...19..716A}
{Akaike} H.,  1974, IEEE Transactions on Automatic Control, \href
  {https://ui.adsabs.harvard.edu/abs/1974ITAC...19..716A} {19, 716}

\bibitem[\protect\citeauthoryear{{Aloisio}, {Berezinsky}, {Blasi}, {Gazizov},
  {Grigorieva}  \& {Hnatyk}}{{Aloisio} et~al.}{2007}]{Aloisio2007}
{Aloisio} R.,  {Berezinsky} V.,  {Blasi} P.,  {Gazizov} A.,  {Grigorieva} S.,
  {Hnatyk} B.,  2007, \mn@doi [Astroparticle Physics]
  {10.1016/j.astropartphys.2006.09.004}, \href
  {https://ui.adsabs.harvard.edu/abs/2007APh....27...76A} {27, 76}

\bibitem[\protect\citeauthoryear{{Astiasarain}, {Tibaldo}, {Martin},
  {Kn{\"o}dlseder}  \& {Remy}}{{Astiasarain}
  et~al.}{2023}]{Astiasarain2023A&A...671A..47A}
{Astiasarain} X.,  {Tibaldo} L.,  {Martin} P.,  {Kn{\"o}dlseder} J.,   {Remy}
  Q.,  2023, \mn@doi [\aap] {10.1051/0004-6361/202245573}, \href
  {https://ui.adsabs.harvard.edu/abs/2023A&A...671A..47A} {671, A47}

\bibitem[\protect\citeauthoryear{{Bell}, {Schure}, {Reville}  \&
  {Giacinti}}{{Bell} et~al.}{2013}]{Bell2013}
{Bell} A.~R.,  {Schure} K.~M.,  {Reville} B.,   {Giacinti} G.,  2013, \mn@doi
  [\mnras] {10.1093/mnras/stt179}, \href
  {https://ui.adsabs.harvard.edu/abs/2013MNRAS.431..415B} {431, 415}

\bibitem[\protect\citeauthoryear{Bhadra, Gupta, Nath  \& Sharma}{Bhadra
  et~al.}{2022}]{Bhadra10.1093/mnras/stac023}
Bhadra S.,  Gupta S.,  Nath B.~B.,   Sharma P.,  2022, \mn@doi [Monthly Notices
  of the Royal Astronomical Society] {10.1093/mnras/stac023}, 510, 5579

\bibitem[\protect\citeauthoryear{{Bik} et~al.,}{{Bik} et~al.}{2012}]{Bik2012}
{Bik} A.,  et~al., 2012, \mn@doi [\apj] {10.1088/0004-637X/744/2/87}, \href
  {https://ui.adsabs.harvard.edu/abs/2012ApJ...744...87B} {744, 87}

\bibitem[\protect\citeauthoryear{{Bolatto}, {Wolfire}  \& {Leroy}}{{Bolatto}
  et~al.}{2013}]{Bolatto2013}
{Bolatto} A.~D.,  {Wolfire} M.,   {Leroy} A.~K.,  2013, \mn@doi [\araa]
  {10.1146/annurev-astro-082812-140944}, \href
  {https://ui.adsabs.harvard.edu/abs/2013ARA&A..51..207B} {51, 207}

\bibitem[\protect\citeauthoryear{{Boumis} et~al.,}{{Boumis}
  et~al.}{2022}]{Boumis2022}
{Boumis} P.,  et~al., 2022, \mn@doi [\mnras] {10.1093/mnras/stac412}, \href
  {https://ui.adsabs.harvard.edu/abs/2022MNRAS.512.1658B} {512, 1658}

\bibitem[\protect\citeauthoryear{{Brand} \& {Blitz}}{{Brand} \&
  {Blitz}}{1993}]{Brand1993A&A...275...67B}
{Brand} J.,  {Blitz} L.,  1993, \aap, \href
  {https://ui.adsabs.harvard.edu/abs/1993A&A...275...67B} {275, 67}

\bibitem[\protect\citeauthoryear{{Bruel}}{{Bruel}}{2021}]{Bruel2021A&A...656A..81B}
{Bruel} P.,  2021, \mn@doi [\aap] {10.1051/0004-6361/202141553}, \href
  {https://ui.adsabs.harvard.edu/abs/2021A&A...656A..81B} {656, A81}

\bibitem[\protect\citeauthoryear{{Case} \& {Bhattacharya}}{{Case} \&
  {Bhattacharya}}{1998}]{Case1998ApJ...504..761C}
{Case} G.~L.,  {Bhattacharya} D.,  1998, \mn@doi [\apj] {10.1086/306089}, \href
  {https://ui.adsabs.harvard.edu/abs/1998ApJ...504..761C} {504, 761}

\bibitem[\protect\citeauthoryear{{Crutcher}}{{Crutcher}}{2012}]{Crutcher2012ARA&A..50...29C}
{Crutcher} R.~M.,  2012, \mn@doi [\araa] {10.1146/annurev-astro-081811-125514},
  \href {https://ui.adsabs.harvard.edu/abs/2012ARA&A..50...29C} {50, 29}

\bibitem[\protect\citeauthoryear{{Dame}, {Hartmann}  \& {Thaddeus}}{{Dame}
  et~al.}{2001}]{Dame2001}
{Dame} T.~M.,  {Hartmann} D.,   {Thaddeus} P.,  2001, \mn@doi [\apj]
  {10.1086/318388}, \href
  {https://ui.adsabs.harvard.edu/abs/2001ApJ...547..792D} {547, 792}

\bibitem[\protect\citeauthoryear{{Ezoe}, {Kokubun}, {Makishima}, {Sekimoto}  \&
  {Matsuzaki}}{{Ezoe} et~al.}{2006}]{Ezoe2006}
{Ezoe} Y.,  {Kokubun} M.,  {Makishima} K.,  {Sekimoto} Y.,   {Matsuzaki} K.,
  2006, \mn@doi [\apj] {10.1086/499120}, \href
  {https://ui.adsabs.harvard.edu/abs/2006ApJ...638..860E} {638, 860}

\bibitem[\protect\citeauthoryear{{Feigelson} \& {Townsley}}{{Feigelson} \&
  {Townsley}}{2008}]{Feigelson2008}
{Feigelson} E.~D.,  {Townsley} L.~K.,  2008, \mn@doi [\apj] {10.1086/524031},
  \href {https://ui.adsabs.harvard.edu/abs/2008ApJ...673..354F} {673, 354}

\bibitem[\protect\citeauthoryear{{Finkbeiner}}{{Finkbeiner}}{2003}]{Finkbeiner2003}
{Finkbeiner} D.~P.,  2003, \mn@doi [\apjs] {10.1086/374411}, \href
  {https://ui.adsabs.harvard.edu/abs/2003ApJS..146..407F} {146, 407}

\bibitem[\protect\citeauthoryear{{Foreman-Mackey}, {Hogg}, {Lang}  \&
  {Goodman}}{{Foreman-Mackey} et~al.}{2013}]{Foreman2013}
{Foreman-Mackey} D.,  {Hogg} D.~W.,  {Lang} D.,   {Goodman} J.,  2013, \mn@doi
  [\pasp] {10.1086/670067}, \href
  {https://ui.adsabs.harvard.edu/abs/2013PASP..125..306F} {125, 306}

\bibitem[\protect\citeauthoryear{{Ge}, {Sun}, {Yang}, {Liang}  \& {Liang}}{{Ge}
  et~al.}{2022}]{Ge2022}
{Ge} T.-T.,  {Sun} X.-N.,  {Yang} R.-Z.,  {Liang} Y.-F.,   {Liang} E.-W.,
  2022, \mn@doi [\mnras] {10.1093/mnras/stac2885}, \href
  {https://ui.adsabs.harvard.edu/abs/2022MNRAS.517.5121G} {517, 5121}

\bibitem[\protect\citeauthoryear{{Green}}{{Green}}{2014}]{Green2014}
{Green} D.~A.,  2014, \mn@doi [Bulletin of the Astronomical Society of India]
  {10.48550/arXiv.1409.0637}, \href
  {https://ui.adsabs.harvard.edu/abs/2014BASI...42...47G} {42, 47}

\bibitem[\protect\citeauthoryear{{Gupta}, {Nath}  \& {Sharma}}{{Gupta}
  et~al.}{2018}]{Gupta2018MNRAS.479.5220G}
{Gupta} S.,  {Nath} B.~B.,   {Sharma} P.,  2018, \mn@doi [\mnras]
  {10.1093/mnras/sty1846}, \href
  {https://ui.adsabs.harvard.edu/abs/2018MNRAS.479.5220G} {479, 5220}

\bibitem[\protect\citeauthoryear{{Gupta}, {Nath}, {Sharma}  \&
  {Eichler}}{{Gupta} et~al.}{2020}]{Gupta2020MNRAS.493.3159G}
{Gupta} S.,  {Nath} B.~B.,  {Sharma} P.,   {Eichler} D.,  2020, \mn@doi
  [\mnras] {10.1093/mnras/staa286}, \href
  {https://ui.adsabs.harvard.edu/abs/2020MNRAS.493.3159G} {493, 3159}

\bibitem[\protect\citeauthoryear{{H.~E.~S.~S. Collaboration}
  et~al.,}{{H.~E.~S.~S. Collaboration}
  et~al.}{2015}]{30DorC2015Sci...347..406H}
{H.~E.~S.~S. Collaboration} et~al., 2015, \mn@doi [Science]
  {10.1126/science.1261313}, \href
  {https://ui.adsabs.harvard.edu/abs/2015Sci...347..406H} {347, 406}

\bibitem[\protect\citeauthoryear{{HI4PI Collaboration} et~al.,}{{HI4PI
  Collaboration} et~al.}{2016}]{HI4PI2016}
{HI4PI Collaboration} et~al., 2016, \mn@doi [\aap]
  {10.1051/0004-6361/201629178}, \href
  {https://ui.adsabs.harvard.edu/abs/2016A&A...594A.116H} {594, A116}

\bibitem[\protect\citeauthoryear{{Hachisuka} et~al.,}{{Hachisuka}
  et~al.}{2006}]{Hachisuka2006}
{Hachisuka} K.,  et~al., 2006, \mn@doi [\apj] {10.1086/502962}, \href
  {https://ui.adsabs.harvard.edu/abs/2006ApJ...645..337H} {645, 337}

\bibitem[\protect\citeauthoryear{{Kafexhiu}, {Aharonian}, {Taylor}  \&
  {Vila}}{{Kafexhiu} et~al.}{2014}]{Kafexhiu2014}
{Kafexhiu} E.,  {Aharonian} F.,  {Taylor} A.~M.,   {Vila} G.~S.,  2014, \mn@doi
  [\prd] {10.1103/PhysRevD.90.123014}, \href
  {https://ui.adsabs.harvard.edu/abs/2014PhRvD..90l3014K} {90, 123014}

\bibitem[\protect\citeauthoryear{{Katagiri}, {Yoshida}, {Ballet}, {Grondin},
  {Hanabata}, {Hewitt}, {Kubo}  \& {Lemoine-Goumard}}{{Katagiri}
  et~al.}{2016}]{Katagiri2016}
{Katagiri} H.,  {Yoshida} K.,  {Ballet} J.,  {Grondin} M.~H.,  {Hanabata} Y.,
  {Hewitt} J.~W.,  {Kubo} H.,   {Lemoine-Goumard} M.,  2016, \mn@doi [\apj]
  {10.3847/0004-637X/818/2/114}, \href
  {https://ui.adsabs.harvard.edu/abs/2016ApJ...818..114K} {818, 114}

\bibitem[\protect\citeauthoryear{{Khangulyan}, {Aharonian}  \&
  {Kelner}}{{Khangulyan} et~al.}{2014}]{Khangulyan2014}
{Khangulyan} D.,  {Aharonian} F.~A.,   {Kelner} S.~R.,  2014, \mn@doi [\apj]
  {10.1088/0004-637X/783/2/100}, \href
  {https://ui.adsabs.harvard.edu/abs/2014ApJ...783..100K} {783, 100}

\bibitem[\protect\citeauthoryear{{Kiminki}, {Kim}, {Bagley}, {Sherry}  \&
  {Rieke}}{{Kiminki} et~al.}{2015}]{Kiminki2015ApJ...813...42K}
{Kiminki} M.~M.,  {Kim} J.~S.,  {Bagley} M.~B.,  {Sherry} W.~H.,   {Rieke}
  G.~H.,  2015, \mn@doi [\apj] {10.1088/0004-637X/813/1/42}, \href
  {https://ui.adsabs.harvard.edu/abs/2015ApJ...813...42K} {813, 42}

\bibitem[\protect\citeauthoryear{{Lagage} \& {Cesarsky}}{{Lagage} \&
  {Cesarsky}}{1983}]{Lagage1983}
{Lagage} P.~O.,  {Cesarsky} C.~J.,  1983, \aap, \href
  {https://ui.adsabs.harvard.edu/abs/1983A&A...125..249L} {125, 249}

\bibitem[\protect\citeauthoryear{{Lande} et~al.,}{{Lande}
  et~al.}{2012}]{Lande2012}
{Lande} J.,  et~al., 2012, \mn@doi [\apj] {10.1088/0004-637X/756/1/5}, \href
  {https://ui.adsabs.harvard.edu/abs/2012ApJ...756....5L} {756, 5}

\bibitem[\protect\citeauthoryear{{Lazendic} \& {Slane}}{{Lazendic} \&
  {Slane}}{2006}]{Lazendic2006}
{Lazendic} J.~S.,  {Slane} P.~O.,  2006, \mn@doi [\apj] {10.1086/505380}, \href
  {https://ui.adsabs.harvard.edu/abs/2006ApJ...647..350L} {647, 350}

\bibitem[\protect\citeauthoryear{{Lebrun} et~al.,}{{Lebrun}
  et~al.}{1983}]{Lebrun1983}
{Lebrun} F.,  et~al., 1983, \mn@doi [\apj] {10.1086/161440}, \href
  {https://ui.adsabs.harvard.edu/abs/1983ApJ...274..231L} {274, 231}

\bibitem[\protect\citeauthoryear{{Liu}, {Yang}  \& {Chen}}{{Liu}
  et~al.}{2022}]{Liu2022}
{Liu} B.,  {Yang} R.-z.,   {Chen} Z.,  2022, \mn@doi [\mnras]
  {10.1093/mnras/stac1252}, \href
  {https://ui.adsabs.harvard.edu/abs/2022MNRAS.513.4747L} {513, 4747}

\bibitem[\protect\citeauthoryear{{Lorimer}, {Lyne}  \& {Camilo}}{{Lorimer}
  et~al.}{1998}]{Lorimer1998}
{Lorimer} D.~R.,  {Lyne} A.~G.,   {Camilo} F.,  1998, \mn@doi [\aap]
  {10.48550/arXiv.astro-ph/9801096}, \href
  {https://ui.adsabs.harvard.edu/abs/1998A&A...331.1002L} {331, 1002}

\bibitem[\protect\citeauthoryear{{Lorimer} et~al.,}{{Lorimer}
  et~al.}{2006}]{Lorimer2006MNRAS.372..777L}
{Lorimer} D.~R.,  et~al., 2006, \mn@doi [\mnras]
  {10.1111/j.1365-2966.2006.10887.x}, \href
  {https://ui.adsabs.harvard.edu/abs/2006MNRAS.372..777L} {372, 777}

\bibitem[\protect\citeauthoryear{{Massey}, {Johnson}  \&
  {Degioia-Eastwood}}{{Massey} et~al.}{1995}]{Massey1995ApJ...454..151M}
{Massey} P.,  {Johnson} K.~E.,   {Degioia-Eastwood} K.,  1995, \mn@doi [\apj]
  {10.1086/176474}, \href
  {https://ui.adsabs.harvard.edu/abs/1995ApJ...454..151M} {454, 151}

\bibitem[\protect\citeauthoryear{{Mathys}}{{Mathys}}{1989}]{Mathys1989}
{Mathys} G.,  1989, \aaps, \href
  {https://ui.adsabs.harvard.edu/abs/1989A&AS...81..237M} {81, 237}

\bibitem[\protect\citeauthoryear{{Navarete}, {Figueredo}, {Damineli},
  {Mois{\'e}s}, {Blum}  \& {Conti}}{{Navarete} et~al.}{2011}]{Navarete2011}
{Navarete} F.,  {Figueredo} E.,  {Damineli} A.,  {Mois{\'e}s} A.~P.,  {Blum}
  R.~D.,   {Conti} P.~S.,  2011, \mn@doi [\aj] {10.1088/0004-6256/142/3/67},
  \href {https://ui.adsabs.harvard.edu/abs/2011AJ....142...67N} {142, 67}

\bibitem[\protect\citeauthoryear{{Oey}, {Watson}, {Kern}  \& {Walth}}{{Oey}
  et~al.}{2005}]{Oey2005}
{Oey} M.~S.,  {Watson} A.~M.,  {Kern} K.,   {Walth} G.~L.,  2005, \mn@doi [\aj]
  {10.1086/426333}, \href
  {https://ui.adsabs.harvard.edu/abs/2005AJ....129..393O} {129, 393}

\bibitem[\protect\citeauthoryear{{Parizot}, {Marcowith}, {van der Swaluw},
  {Bykov}  \& {Tatischeff}}{{Parizot}
  et~al.}{2004}]{Parizot2004A&A...424..747P}
{Parizot} E.,  {Marcowith} A.,  {van der Swaluw} E.,  {Bykov} A.~M.,
  {Tatischeff} V.,  2004, \mn@doi [\aap] {10.1051/0004-6361:20041269}, \href
  {https://ui.adsabs.harvard.edu/abs/2004A&A...424..747P} {424, 747}

\bibitem[\protect\citeauthoryear{{Peron}, {Casanova}, {Gabici}, {Baghmanyan}
  \& {Aharonian}}{{Peron} et~al.}{2024}]{Peron2024NatAs.tmp...10P}
{Peron} G.,  {Casanova} S.,  {Gabici} S.,  {Baghmanyan} V.,   {Aharonian} F.,
  2024, \mn@doi [Nature Astronomy] {10.1038/s41550-023-02168-6}, \href
  {https://ui.adsabs.harvard.edu/abs/2024NatAs.tmp...10P} {}

\bibitem[\protect\citeauthoryear{{Planck Collaboration} et~al.,}{{Planck
  Collaboration} et~al.}{2011}]{Planck2011A&A...536A..13P}
{Planck Collaboration} et~al., 2011, \mn@doi [\aap]
  {10.1051/0004-6361/201116471}, \href
  {https://ui.adsabs.harvard.edu/abs/2011A&A...536A..13P} {536, A13}

\bibitem[\protect\citeauthoryear{{Planck Collaboration} et~al.,}{{Planck
  Collaboration} et~al.}{2016}]{Planck2016}
{Planck Collaboration} et~al., 2016, \mn@doi [\aap]
  {10.1051/0004-6361/201525967}, \href
  {https://ui.adsabs.harvard.edu/abs/2016A&A...594A..10P} {594, A10}

\bibitem[\protect\citeauthoryear{{Polychroni}, {Moore}  \&
  {Allsopp}}{{Polychroni} et~al.}{2012}]{Polychroni2012MNRAS.422.2992P}
{Polychroni} D.,  {Moore} T.~J.~T.,   {Allsopp} J.,  2012, \mn@doi [\mnras]
  {10.1111/j.1365-2966.2012.20803.x}, \href
  {https://ui.adsabs.harvard.edu/abs/2012MNRAS.422.2992P} {422, 2992}

\bibitem[\protect\citeauthoryear{{Popescu}, {Yang}, {Tuffs}, {Natale},
  {Rushton}  \& {Aharonian}}{{Popescu}
  et~al.}{2017}]{Popescu2017MNRAS.470.2539P}
{Popescu} C.~C.,  {Yang} R.,  {Tuffs} R.~J.,  {Natale} G.,  {Rushton} M.,
  {Aharonian} F.,  2017, \mn@doi [\mnras] {10.1093/mnras/stx1282}, \href
  {https://ui.adsabs.harvard.edu/abs/2017MNRAS.470.2539P} {470, 2539}

\bibitem[\protect\citeauthoryear{{Portegies Zwart}, {McMillan}  \&
  {Gieles}}{{Portegies Zwart} et~al.}{2010}]{Portegies2010ARA&A..48..431P}
{Portegies Zwart} S.~F.,  {McMillan} S. L.~W.,   {Gieles} M.,  2010, \mn@doi
  [\araa] {10.1146/annurev-astro-081309-130834}, \href
  {https://ui.adsabs.harvard.edu/abs/2010ARA&A..48..431P} {48, 431}

\bibitem[\protect\citeauthoryear{{Protassov}, {van Dyk}, {Connors}, {Kashyap}
  \& {Siemiginowska}}{{Protassov} et~al.}{2002}]{Protassov2002ApJ...571..545P}
{Protassov} R.,  {van Dyk} D.~A.,  {Connors} A.,  {Kashyap} V.~L.,
  {Siemiginowska} A.,  2002, \mn@doi [\apj] {10.1086/339856}, \href
  {https://ui.adsabs.harvard.edu/abs/2002ApJ...571..545P} {571, 545}

\bibitem[\protect\citeauthoryear{{Reid}, {Dame}, {Menten}  \&
  {Brunthaler}}{{Reid} et~al.}{2016}]{Reid2016}
{Reid} M.~J.,  {Dame} T.~M.,  {Menten} K.~M.,   {Brunthaler} A.,  2016, \mn@doi
  [\apj] {10.3847/0004-637X/823/2/77}, \href
  {https://ui.adsabs.harvard.edu/abs/2016ApJ...823...77R} {823, 77}

\bibitem[\protect\citeauthoryear{{Rom{\'a}n-Z{\'u}{\~n}iga}, {Ybarra},
  {Meg{\'\i}as}, {Tapia}, {Lada}  \& {Alves}}{{Rom{\'a}n-Z{\'u}{\~n}iga}
  et~al.}{2015}]{Román-Zúñiga2015}
{Rom{\'a}n-Z{\'u}{\~n}iga} C.~G.,  {Ybarra} J.~E.,  {Meg{\'\i}as} G.~D.,
  {Tapia} M.,  {Lada} E.~A.,   {Alves} J.~F.,  2015, \mn@doi [\aj]
  {10.1088/0004-6256/150/3/80}, \href
  {https://ui.adsabs.harvard.edu/abs/2015AJ....150...80R} {150, 80}

\bibitem[\protect\citeauthoryear{{Routledge}, {Dewdney}, {Landecker}  \&
  {Vaneldik}}{{Routledge} et~al.}{1991}]{Routledge1991}
{Routledge} D.,  {Dewdney} P.~E.,  {Landecker} T.~L.,   {Vaneldik} J.~F.,
  1991, \aap, \href {https://ui.adsabs.harvard.edu/abs/1991A&A...247..529R}
  {247, 529}

\bibitem[\protect\citeauthoryear{{Russeil}}{{Russeil}}{2003}]{Russeil2003A&A...397..133R}
{Russeil} D.,  2003, \mn@doi [\aap] {10.1051/0004-6361:20021504}, \href
  {https://ui.adsabs.harvard.edu/abs/2003A&A...397..133R} {397, 133}

\bibitem[\protect\citeauthoryear{{Sodroski}, {Odegard}, {Arendt}, {Dwek},
  {Weiland}, {Hauser}  \& {Kelsall}}{{Sodroski} et~al.}{1997}]{Sodroski1997}
{Sodroski} T.~J.,  {Odegard} N.,  {Arendt} R.~G.,  {Dwek} E.,  {Weiland} J.~L.,
   {Hauser} M.~G.,   {Kelsall} T.,  1997, \mn@doi [\apj] {10.1086/303961},
  \href {https://ui.adsabs.harvard.edu/abs/1997ApJ...480..173S} {480, 173}

\bibitem[\protect\citeauthoryear{{Strong} \& {Moskalenko}}{{Strong} \&
  {Moskalenko}}{1998}]{Strong1998ApJ...509..212S}
{Strong} A.~W.,  {Moskalenko} I.~V.,  1998, \mn@doi [\apj] {10.1086/306470},
  \href {https://ui.adsabs.harvard.edu/abs/1998ApJ...509..212S} {509, 212}

\bibitem[\protect\citeauthoryear{{Sun}, {Yang}  \& {Wang}}{{Sun}
  et~al.}{2020a}]{sunRSGC1}
{Sun} X.-N.,  {Yang} R.-Z.,   {Wang} X.-Y.,  2020a, \mn@doi [\mnras]
  {10.1093/mnras/staa947}, \href
  {https://ui.adsabs.harvard.edu/abs/2020MNRAS.494.3405S} {494, 3405}

\bibitem[\protect\citeauthoryear{{Sun}, {Yang}, {Liang}, {Peng}, {Zhang},
  {Wang}  \& {Aharonian}}{{Sun} et~al.}{2020b}]{sunw40}
{Sun} X.-N.,  {Yang} R.-Z.,  {Liang} Y.-F.,  {Peng} F.-K.,  {Zhang} H.-M.,
  {Wang} X.-Y.,   {Aharonian} F.,  2020b, \mn@doi [\aap]
  {10.1051/0004-6361/202037580}, \href
  {https://ui.adsabs.harvard.edu/abs/2020A&A...639A..80S} {639, A80}

\bibitem[\protect\citeauthoryear{{Sun}, {Yang}  \& {Liang}}{{Sun}
  et~al.}{2022}]{sun22}
{Sun} X.-N.,  {Yang} R.-Z.,   {Liang} E.-W.,  2022, \mn@doi [\aap]
  {10.1051/0004-6361/202142394}, \href
  {https://ui.adsabs.harvard.edu/abs/2022A&A...659A..83S} {659, A83}

\bibitem[\protect\citeauthoryear{{Westerhout}}{{Westerhout}}{1958}]{Westerhout1958}
{Westerhout} G.,  1958, \bain, \href
  {https://ui.adsabs.harvard.edu/abs/1958BAN....14..215W} {14, 215}

\bibitem[\protect\citeauthoryear{Wilks}{Wilks}{1938}]{Wilks1938AnnMathStatist..9..60}
Wilks S.~S.,  1938, \mn@doi [The Annals of Mathematical Statistics]
  {10.1214/aoms/1177732360}, 9, 60

\bibitem[\protect\citeauthoryear{{Xu}, {Reid}, {Zheng}  \& {Menten}}{{Xu}
  et~al.}{2006}]{Xu2006Sci...311...54X}
{Xu} Y.,  {Reid} M.~J.,  {Zheng} X.~W.,   {Menten} K.~M.,  2006, \mn@doi
  [Science] {10.1126/science.1120914}, \href
  {https://ui.adsabs.harvard.edu/abs/2006Sci...311...54X} {311, 54}

\bibitem[\protect\citeauthoryear{{Yamada} et~al.,}{{Yamada}
  et~al.}{2021}]{Yamada2021}
{Yamada} R.~I.,  et~al., 2021, \mn@doi [arXiv e-prints]
  {10.48550/arXiv.2106.02217}, \href
  {https://ui.adsabs.harvard.edu/abs/2021arXiv210602217Y} {p. arXiv:2106.02217}

\bibitem[\protect\citeauthoryear{{Yang} \& {Aharonian}}{{Yang} \&
  {Aharonian}}{2017}]{Yang2017A&A...600A.107Y}
{Yang} R.-z.,  {Aharonian} F.,  2017, \mn@doi [\aap]
  {10.1051/0004-6361/201630213}, \href
  {https://ui.adsabs.harvard.edu/abs/2017A&A...600A.107Y} {600, A107}

\bibitem[\protect\citeauthoryear{{Yang}, {de O{\~n}a Wilhelmi}  \&
  {Aharonian}}{{Yang} et~al.}{2018}]{Yang2018A&A...611A..77Y}
{Yang} R.-z.,  {de O{\~n}a Wilhelmi} E.,   {Aharonian} F.,  2018, \mn@doi
  [\aap] {10.1051/0004-6361/201732045}, \href
  {https://ui.adsabs.harvard.edu/abs/2018A&A...611A..77Y} {611, A77}

\bibitem[\protect\citeauthoryear{{Yang}, {Li}, {Wilhelmi}, {Cui}, {Liu}  \&
  {Aharonian}}{{Yang} et~al.}{2023}]{Yang2023NatAs...7..351Y}
{Yang} R.-z.,  {Li} G.-X.,  {Wilhelmi} E. d.~O.,  {Cui} Y.-D.,  {Liu} B.,
  {Aharonian} F.,  2023, \mn@doi [Nature Astronomy]
  {10.1038/s41550-022-01868-9}, \href
  {https://ui.adsabs.harvard.edu/abs/2023NatAs...7..351Y} {7, 351}

\bibitem[\protect\citeauthoryear{{Zabalza}}{{Zabalza}}{2015}]{Zabalza2015}
{Zabalza} V.,  2015, in 34th International Cosmic Ray Conference (ICRC2015).
  p.~922 (\mn@eprint {arXiv} {1509.03319}), \mn@doi{10.22323/1.236.0922}

\bibitem[\protect\citeauthoryear{{Zhou}, {Yang}, {Fang}, {Su}, {Sun}  \&
  {Chen}}{{Zhou} et~al.}{2016}]{Zhou2016}
{Zhou} X.,  {Yang} J.,  {Fang} M.,  {Su} Y.,  {Sun} Y.,   {Chen} Y.,  2016,
  \mn@doi [\apj] {10.3847/0004-637X/833/1/4}, \href
  {https://ui.adsabs.harvard.edu/abs/2016ApJ...833....4Z} {833, 4}

\bibitem[\protect\citeauthoryear{{Zuo} et~al.,}{{Zuo} et~al.}{2011}]{Zuo2011}
{Zuo} Y.~X.,  et~al., 2011, Acta Astronomica Sinica, \href
  {https://ui.adsabs.harvard.edu/abs/2011AcASn..52..152Z} {52, 152}

\makeatother
\end{thebibliography}


\begin{thebibliography}{}
\makeatletter
\relax
\def\mn@urlcharsother{\let\do\@makeother \do\$\do\&\do\#\do\^\do\_\do\%\do\~}
\def\mn@doi{\begingroup\mn@urlcharsother \@ifnextchar [ {\mn@doi@}
  {\mn@doi@[]}}
\def\mn@doi@[#1]#2{\def\@tempa{#1}\ifx\@tempa\@empty \href
  {http://dx.doi.org/#2} {doi:#2}\else \href {http://dx.doi.org/#2} {#1}\fi
  \endgroup}
\def\mn@eprint#1#2{\mn@eprint@#1:#2::\@nil}
\def\mn@eprint@arXiv#1{\href {http://arxiv.org/abs/#1} {{\tt arXiv:#1}}}
\def\mn@eprint@dblp#1{\href {http://dblp.uni-trier.de/rec/bibtex/#1.xml}
  {dblp:#1}}
\def\mn@eprint@#1:#2:#3:#4\@nil{\def\@tempa {#1}\def\@tempb {#2}\def\@tempc
  {#3}\ifx \@tempc \@empty \let \@tempc \@tempb \let \@tempb \@tempa \fi \ifx
  \@tempb \@empty \def\@tempb {arXiv}\fi \@ifundefined
  {mn@eprint@\@tempb}{\@tempb:\@tempc}{\expandafter \expandafter \csname
  mn@eprint@\@tempb\endcsname \expandafter{\@tempc}}}

\bibitem[\protect\citeauthoryear{{Abdo} et~al.,}{{Abdo} et~al.}{2009a}]{Abdo09}
{Abdo} A.~A.,  et~al., 2009a, \mn@doi [\apjs] {10.1088/0067-0049/183/1/46},
  \href {https://ui.adsabs.harvard.edu/abs/2009ApJS..183...46A} {183, 46}

\bibitem[\protect\citeauthoryear{{Abdo} et~al.,}{{Abdo}
  et~al.}{2009b}]{Abdo09a}
{Abdo} A.~A.,  et~al., 2009b, \mn@doi [\apjl] {10.1088/0004-637X/706/1/L1},
  \href {https://ui.adsabs.harvard.edu/abs/2009ApJ...706L...1A} {706, L1}

\bibitem[\protect\citeauthoryear{{Abdollahi} et~al.,}{{Abdollahi}
  et~al.}{2020}]{Abdollahi20}
{Abdollahi} S.,  et~al., 2020, \mn@doi [\apjs] {10.3847/1538-4365/ab6bcb},
  \href {https://ui.adsabs.harvard.edu/abs/2020ApJS..247...33A} {247, 33}

\bibitem[\protect\citeauthoryear{{Abeysekara} et~al.,}{{Abeysekara}
  et~al.}{2021}]{Abeysekara21}
{Abeysekara} A.~U.,  et~al., 2021, \mn@doi [Nature Astronomy]
  {10.1038/s41550-021-01318-y}, \href
  {https://ui.adsabs.harvard.edu/abs/2021NatAs...5..465A} {5, 465}

\bibitem[\protect\citeauthoryear{{Abramowski} et~al.,}{{Abramowski}
  et~al.}{2012}]{Abramowski12}
{Abramowski} A.,  et~al., 2012, \mn@doi [\aap] {10.1051/0004-6361/201117928},
  \href {https://ui.adsabs.harvard.edu/abs/2012A%26A...537A.114A} {537, A114}

\bibitem[\protect\citeauthoryear{{Ackermann} et~al.,}{{Ackermann}
  et~al.}{2011}]{Ackermann11}
{Ackermann} M.,  et~al., 2011, \mn@doi [Science] {10.1126/science.1210311},
  \href {https://ui.adsabs.harvard.edu/abs/2011Sci...334.1103A} {334, 1103}

\bibitem[\protect\citeauthoryear{{Ackermann} et~al.,}{{Ackermann}
  et~al.}{2017}]{Ackermann17}
{Ackermann} M.,  et~al., 2017, \mn@doi [\apj] {10.3847/1538-4357/aa775a}, \href
  {https://ui.adsabs.harvard.edu/abs/2017ApJ...843..139A} {843, 139}

\bibitem[\protect\citeauthoryear{{Aguilar} et~al.,}{{Aguilar}
  et~al.}{2015}]{Aguilar15}
{Aguilar} M.,  et~al., 2015, \mn@doi [\prl] {10.1103/PhysRevLett.114.171103},
  \href {https://ui.adsabs.harvard.edu/abs/2015PhRvL.114q1103A} {114, 171103}

\bibitem[\protect\citeauthoryear{{Aharonian} et~al.,}{{Aharonian}
  et~al.}{2007}]{Aharonian07}
{Aharonian} F.,  et~al., 2007, \mn@doi [\aap] {10.1051/0004-6361:20066950},
  \href {https://ui.adsabs.harvard.edu/abs/2007A&A...467.1075A} {467, 1075}

\bibitem[\protect\citeauthoryear{{Aharonian}, {Yang}  \& {de O{\~n}a
  Wilhelmi}}{{Aharonian} et~al.}{2019}]{Aharonian19}
{Aharonian} F.,  {Yang} R.,   {de O{\~n}a Wilhelmi} E.,  2019, \mn@doi [Nature
  Astronomy] {10.1038/s41550-019-0724-0}, \href
  {https://ui.adsabs.harvard.edu/abs/2019NatAs...3..561A} {3, 561}

\bibitem[\protect\citeauthoryear{{Aharonian} et~al.,}{{Aharonian}
  et~al.}{2022}]{Aharonian22}
{Aharonian} F.,  et~al., 2022, arXiv e-prints, \href
  {https://ui.adsabs.harvard.edu/abs/2022arXiv220710921A} {p. arXiv:2207.10921}

\bibitem[\protect\citeauthoryear{{Akaike}}{{Akaike}}{1974}]{Akaike1974}
{Akaike} H.,  1974, IEEE Transactions on Automatic Control, \href
  {https://ui.adsabs.harvard.edu/abs/1974ITAC...19..716A} {19, 716}

\bibitem[\protect\citeauthoryear{{Ascenso}, {Alves}, {Vicente}  \&
  {Lago}}{{Ascenso} et~al.}{2007}]{Ascenso07}
{Ascenso} J.,  {Alves} J.,  {Vicente} S.,   {Lago} M.~T.~V.~T.,  2007, \mn@doi
  [\aap] {10.1051/0004-6361:20077210}, \href
  {https://ui.adsabs.harvard.edu/abs/2007A&A...476..199A} {476, 199}

\bibitem[\protect\citeauthoryear{{Baade} \& {Zwicky}}{{Baade} \&
  {Zwicky}}{1934}]{1934SNRs}
{Baade} W.,  {Zwicky} F.,  1934, \mn@doi [Proceedings of the National Academy
  of Science] {10.1073/pnas.20.5.259}, \href
  {https://ui.adsabs.harvard.edu/abs/1934PNAS...20..259B} {20, 259}

\bibitem[\protect\citeauthoryear{{Balbo} \& {Walter}}{{Balbo} \&
  {Walter}}{2017}]{Balbo17}
{Balbo} M.,  {Walter} R.,  2017, \mn@doi [\aap] {10.1051/0004-6361/201629640},
  \href {https://ui.adsabs.harvard.edu/abs/2017A&A...603A.111B} {603, A111}

\bibitem[\protect\citeauthoryear{{Ballet}, {Burnett}, {Digel}  \&
  {Lott}}{{Ballet} et~al.}{2020}]{Ballet20}
{Ballet} J.,  {Burnett} T.~H.,  {Digel} S.~W.,   {Lott} B.,  2020, arXiv
  e-prints, \href {https://ui.adsabs.harvard.edu/abs/2020arXiv200511208B} {p.
  arXiv:2005.11208}

\bibitem[\protect\citeauthoryear{{Bisht}, {Zhu}, {Yadav}, {Ganesh}, {Rangwal},
  {Durgapal}, {Sariya}  \& {Jiang}}{{Bisht} et~al.}{2021}]{Bisht21}
{Bisht} D.,  {Zhu} Q.,  {Yadav} R.~K.~S.,  {Ganesh} S.,  {Rangwal} G.,
  {Durgapal} A.,  {Sariya} D.~P.,   {Jiang} I.-G.,  2021, \mn@doi [\mnras]
  {10.1093/mnras/stab691}, \href
  {https://ui.adsabs.harvard.edu/abs/2021MNRAS.503.5929B} {503, 5929}

\bibitem[\protect\citeauthoryear{{Bolatto}, {Wolfire}  \& {Leroy}}{{Bolatto}
  et~al.}{2013}]{Bolatto13}
{Bolatto} A.~D.,  {Wolfire} M.,   {Leroy} A.~K.,  2013, \mn@doi [\araa]
  {10.1146/annurev-astro-082812-140944}, \href
  {https://ui.adsabs.harvard.edu/abs/2013ARA&A..51..207B} {51, 207}

\bibitem[\protect\citeauthoryear{{Bykov}, {Ellison}, {Gladilin}  \&
  {Osipov}}{{Bykov} et~al.}{2015}]{bykov15}
{Bykov} A.~M.,  {Ellison} D.~C.,  {Gladilin} P.~E.,   {Osipov} S.~M.,  2015,
  \mn@doi [\mnras] {10.1093/mnras/stv1606}, \href
  {https://ui.adsabs.harvard.edu/abs/2015MNRAS.453..113B} {453, 113}

\bibitem[\protect\citeauthoryear{{Dame}, {Hartmann}  \& {Thaddeus}}{{Dame}
  et~al.}{2001}]{Dame01}
{Dame} T.~M.,  {Hartmann} D.,   {Thaddeus} P.,  2001, \mn@doi [\apj]
  {10.1086/318388}, \href
  {https://ui.adsabs.harvard.edu/abs/2001ApJ...547..792D} {547, 792}

\bibitem[\protect\citeauthoryear{{Damineli} et~al.,}{{Damineli}
  et~al.}{2008}]{Damineli08}
{Damineli} A.,  et~al., 2008, \mn@doi [\mnras]
  {10.1111/j.1365-2966.2007.12815.x}, \href
  {https://ui.adsabs.harvard.edu/abs/2008MNRAS.384.1649D} {384, 1649}

\bibitem[\protect\citeauthoryear{{Danilenko}, {Kirichenko}, {Sollerman},
  {Shibanov}  \& {Zyuzin}}{{Danilenko} et~al.}{2013}]{Danilenko13}
{Danilenko} A.,  {Kirichenko} A.,  {Sollerman} J.,  {Shibanov} Y.,   {Zyuzin}
  D.,  2013, \mn@doi [\aap] {10.1051/0004-6361/201220161}, \href
  {https://ui.adsabs.harvard.edu/abs/2013A&A...552A.127D} {552, A127}

\bibitem[\protect\citeauthoryear{{De Becker} \& {Raucq}}{{De Becker} \&
  {Raucq}}{2013}]{DeBecker13}
{De Becker} M.,  {Raucq} F.,  2013, \mn@doi [\aap]
  {10.1051/0004-6361/201322074}, \href
  {https://ui.adsabs.harvard.edu/abs/2013A&A...558A..28D} {558, A28}

\bibitem[\protect\citeauthoryear{{Del Valle} \& {Romero}}{{Del Valle} \&
  {Romero}}{2012}]{delValle12}
{Del Valle} M.~V.,  {Romero} G.~E.,  2012, \mn@doi [\aap]
  {10.1051/0004-6361/201218937}, \href
  {https://ui.adsabs.harvard.edu/abs/2012A&A...543A..56D} {543, A56}

\bibitem[\protect\citeauthoryear{{Ezoe}, {Kokubun}, {Makishima}, {Sekimoto}  \&
  {Matsuzaki}}{{Ezoe} et~al.}{2006}]{Ezoe06}
{Ezoe} Y.,  {Kokubun} M.,  {Makishima} K.,  {Sekimoto} Y.,   {Matsuzaki} K.,
  2006, \mn@doi [\apj] {10.1086/499120}, \href
  {https://ui.adsabs.harvard.edu/abs/2006ApJ...638..860E} {638, 860}

\bibitem[\protect\citeauthoryear{{Farnier}, {Walter}  \& {Leyder}}{{Farnier}
  et~al.}{2011}]{Farnier11}
{Farnier} C.,  {Walter} R.,   {Leyder} J.~C.,  2011, \mn@doi [\aap]
  {10.1051/0004-6361/201015590}, \href
  {https://ui.adsabs.harvard.edu/abs/2011A&A...526A..57F} {526, A57}

\bibitem[\protect\citeauthoryear{{Finkbeiner}}{{Finkbeiner}}{2003}]{Finkbeiner03}
{Finkbeiner} D.~P.,  2003, \mn@doi [\apjs] {10.1086/374411}, \href
  {https://ui.adsabs.harvard.edu/abs/2003ApJS..146..407F} {146, 407}

\bibitem[\protect\citeauthoryear{{Fujita} et~al.,}{{Fujita}
  et~al.}{2021}]{Fujita21}
{Fujita} S.,  et~al., 2021, \mn@doi [\pasj] {10.1093/pasj/psaa078}, \href
  {https://ui.adsabs.harvard.edu/abs/2021PASJ...73S.201F} {73, S201}

\bibitem[\protect\citeauthoryear{{G{\"o}ppl} \& {Preibisch}}{{G{\"o}ppl} \&
  {Preibisch}}{2022}]{G22}
{G{\"o}ppl} C.,  {Preibisch} T.,  2022, arXiv e-prints, \href
  {https://ui.adsabs.harvard.edu/abs/2022arXiv220109097G} {p. arXiv:2201.09097}

\bibitem[\protect\citeauthoryear{{Gupta} \& {Razzaque}}{{Gupta} \&
  {Razzaque}}{2017}]{Gupta17}
{Gupta} N.,  {Razzaque} S.,  2017, \mn@doi [\prd] {10.1103/PhysRevD.96.123017},
  \href {https://ui.adsabs.harvard.edu/abs/2017PhRvD..96l3017G} {96, 123017}

\bibitem[\protect\citeauthoryear{{H.~E.~S.~S. Collaboration}
  et~al.,}{{H.~E.~S.~S. Collaboration} et~al.}{2011}]{HESS11}
{H.~E.~S.~S. Collaboration} et~al., 2011, \mn@doi [\aap]
  {10.1051/0004-6361/201015290}, \href
  {https://ui.adsabs.harvard.edu/abs/2011A&A...525A..46H} {525, A46}

\bibitem[\protect\citeauthoryear{{H.~E.~S.~S. Collaboration}
  et~al.,}{{H.~E.~S.~S. Collaboration} et~al.}{2020}]{HESS20}
{H.~E.~S.~S. Collaboration} et~al., 2020, \mn@doi [\aap]
  {10.1051/0004-6361/201936761}, \href
  {https://ui.adsabs.harvard.edu/abs/2020A&A...635A.167H} {635, A167}

\bibitem[\protect\citeauthoryear{{HESS Collaboration} et~al.,}{{HESS
  Collaboration} et~al.}{2012}]{HESS12}
{HESS Collaboration} et~al., 2012, \mn@doi [\mnras]
  {10.1111/j.1365-2966.2012.21180.x}, \href
  {https://ui.adsabs.harvard.edu/abs/2012MNRAS.424..128H} {424, 128}

\bibitem[\protect\citeauthoryear{{H.E.S.S.~Collaboration}
  et~al.,}{{H.E.S.S.~Collaboration} et~al.}{2015}]{Abramowski15}
{H.E.S.S.~Collaboration} et~al., 2015, \mn@doi [Science]
  {10.1126/science.1261313}, \href
  {https://ui.adsabs.harvard.edu/abs/2015Sci...347..406H} {347, 406}

\bibitem[\protect\citeauthoryear{{HI4PI Collaboration} et~al.,}{{HI4PI
  Collaboration} et~al.}{2016}]{HI4PI16}
{HI4PI Collaboration} et~al., 2016, \mn@doi [\aap]
  {10.1051/0004-6361/201629178}, \href
  {https://ui.adsabs.harvard.edu/abs/2016A&A...594A.116H} {594, A116}

\bibitem[\protect\citeauthoryear{{Hamaguchi} et~al.,}{{Hamaguchi}
  et~al.}{2007}]{Hamaguchi07}
{Hamaguchi} K.,  et~al., 2007, \mn@doi [\pasj] {10.1093/pasj/59.sp1.S151},
  \href {https://ui.adsabs.harvard.edu/abs/2007PASJ...59S.151H} {59, 151}

\bibitem[\protect\citeauthoryear{{Hamaguchi} et~al.,}{{Hamaguchi}
  et~al.}{2018}]{Hamaguchi18}
{Hamaguchi} K.,  et~al., 2018, \mn@doi [Nature Astronomy]
  {10.1038/s41550-018-0505-1}, \href
  {https://ui.adsabs.harvard.edu/abs/2018NatAs...2..731H} {2, 731}

\bibitem[\protect\citeauthoryear{{Hillier}, {Davidson}, {Ishibashi}  \&
  {Gull}}{{Hillier} et~al.}{2001}]{Hillier01}
{Hillier} D.~J.,  {Davidson} K.,  {Ishibashi} K.,   {Gull} T.,  2001, \mn@doi
  [\apj] {10.1086/320948}, \href
  {https://ui.adsabs.harvard.edu/abs/2001ApJ...553..837H} {553, 837}

\bibitem[\protect\citeauthoryear{{Huber}, {Tchernin}, {Eckert}, {Farnier},
  {Manalaysay}, {Straumann}  \& {Walter}}{{Huber} et~al.}{2013}]{Huber13}
{Huber} B.,  {Tchernin} C.,  {Eckert} D.,  {Farnier} C.,  {Manalaysay} A.,
  {Straumann} U.,   {Walter} R.,  2013, \mn@doi [\aap]
  {10.1051/0004-6361/201321947}, \href
  {https://ui.adsabs.harvard.edu/abs/2013A&A...560A..64H} {560, A64}

\bibitem[\protect\citeauthoryear{{Jean}, {Cheung}, {Ojha}, {van Zyl}  \&
  {Angioni}}{{Jean} et~al.}{2018}]{Jean18}
{Jean} P.,  {Cheung} C.~C.,  {Ojha} R.,  {van Zyl} P.,   {Angioni} R.,  2018,
  The Astronomer's Telegram, \href
  {https://ui.adsabs.harvard.edu/abs/2018ATel11546....1J} {11546, 1}

\bibitem[\protect\citeauthoryear{{Kafexhiu}, {Aharonian}, {Taylor}  \&
  {Vila}}{{Kafexhiu} et~al.}{2014}]{Kafexhiu14}
{Kafexhiu} E.,  {Aharonian} F.,  {Taylor} A.~M.,   {Vila} G.~S.,  2014, \mn@doi
  [\prd] {10.1103/PhysRevD.90.123014}, \href
  {https://ui.adsabs.harvard.edu/abs/2014PhRvD..90l3014K} {90, 123014}

\bibitem[\protect\citeauthoryear{{Kelner}, {Aharonian}  \& {Bugayov}}{{Kelner}
  et~al.}{2006}]{Kelner06}
{Kelner} S.~R.,  {Aharonian} F.~A.,   {Bugayov} V.~V.,  2006, \mn@doi [\prd]
  {10.1103/PhysRevD.74.034018}, \href
  {https://ui.adsabs.harvard.edu/abs/2006PhRvD..74c4018K} {74, 034018}

\bibitem[\protect\citeauthoryear{{Khangulyan}, {Aharonian}  \&
  {Kelner}}{{Khangulyan} et~al.}{2014}]{Khangulyan14}
{Khangulyan} D.,  {Aharonian} F.~A.,   {Kelner} S.~R.,  2014, \mn@doi [\apj]
  {10.1088/0004-637X/783/2/100}, \href
  {https://ui.adsabs.harvard.edu/abs/2014ApJ...783..100K} {783, 100}

\bibitem[\protect\citeauthoryear{{Lande} et~al.,}{{Lande}
  et~al.}{2012}]{Lande12}
{Lande} J.,  et~al., 2012, \mn@doi [\apj] {10.1088/0004-637X/756/1/5}, \href
  {https://ui.adsabs.harvard.edu/abs/2012ApJ...756....5L} {756, 5}

\bibitem[\protect\citeauthoryear{{Lebrun} et~al.,}{{Lebrun}
  et~al.}{1983}]{Lebrun1983}
{Lebrun} F.,  et~al., 1983, \mn@doi [\apj] {10.1086/161440}, \href
  {https://ui.adsabs.harvard.edu/abs/1983ApJ...274..231L} {274, 231}

\bibitem[\protect\citeauthoryear{{Liu} \& {Yang}}{{Liu} \&
  {Yang}}{2022}]{LiuYang22}
{Liu} B.,  {Yang} R.-z.,  2022, \mn@doi [\aap] {10.1051/0004-6361/202039759},
  \href {https://ui.adsabs.harvard.edu/abs/2022A&A...659A.101L} {659, A101}

\bibitem[\protect\citeauthoryear{{Liu}, {Yang}  \& {Chen}}{{Liu}
  et~al.}{2022}]{Liu22}
{Liu} B.,  {Yang} R.-z.,   {Chen} Z.,  2022, arXiv e-prints, \href
  {https://ui.adsabs.harvard.edu/abs/2022arXiv220506430L} {p. arXiv:2205.06430}

\bibitem[\protect\citeauthoryear{{Mart{\'\i}-Devesa} \&
  {Reimer}}{{Mart{\'\i}-Devesa} \& {Reimer}}{2021}]{Mart21}
{Mart{\'\i}-Devesa} G.,  {Reimer} O.,  2021, \mn@doi [\aap]
  {10.1051/0004-6361/202140451}, \href
  {https://ui.adsabs.harvard.edu/abs/2021A&A...654A..44M} {654, A44}

\bibitem[\protect\citeauthoryear{{Mestre} et~al.,}{{Mestre}
  et~al.}{2021}]{Mestre21}
{Mestre} E.,  et~al., 2021, \mn@doi [\mnras] {10.1093/mnras/stab1455}, \href
  {https://ui.adsabs.harvard.edu/abs/2021MNRAS.505.2731M} {505, 2731}

\bibitem[\protect\citeauthoryear{{Morlino}, {Blasi}, {Peretti}  \&
  {Cristofari}}{{Morlino} et~al.}{2021}]{Morlino21}
{Morlino} G.,  {Blasi} P.,  {Peretti} E.,   {Cristofari} P.,  2021, \mn@doi
  [\mnras] {10.1093/mnras/stab690}, \href
  {https://ui.adsabs.harvard.edu/abs/2021MNRAS.504.6096M} {504, 6096}

\bibitem[\protect\citeauthoryear{{Ohm}, {Zabalza}, {Hinton}  \& {Parkin}}{{Ohm}
  et~al.}{2015}]{Ohm15}
{Ohm} S.,  {Zabalza} V.,  {Hinton} J.~A.,   {Parkin} E.~R.,  2015, \mn@doi
  [\mnras] {10.1093/mnrasl/slv032}, \href
  {https://ui.adsabs.harvard.edu/abs/2015MNRAS.449L.132O} {449, L132}

\bibitem[\protect\citeauthoryear{{Parkin}, {Pittard}, {Corcoran}, {Hamaguchi}
  \& {Stevens}}{{Parkin} et~al.}{2009}]{Parkin09}
{Parkin} E.~R.,  {Pittard} J.~M.,  {Corcoran} M.~F.,  {Hamaguchi} K.,
  {Stevens} I.~R.,  2009, \mn@doi [\mnras] {10.1111/j.1365-2966.2009.14475.x},
  \href {https://ui.adsabs.harvard.edu/abs/2009MNRAS.394.1758P} {394, 1758}

\bibitem[\protect\citeauthoryear{{Pittard} \& {Corcoran}}{{Pittard} \&
  {Corcoran}}{2002}]{Pittard02}
{Pittard} J.~M.,  {Corcoran} M.~F.,  2002, \mn@doi [\aap]
  {10.1051/0004-6361:20020025}, \href
  {https://ui.adsabs.harvard.edu/abs/2002A&A...383..636P} {383, 636}

\bibitem[\protect\citeauthoryear{{Planck Collaboration} et~al.,}{{Planck
  Collaboration} et~al.}{2016}]{Planck16}
{Planck Collaboration} et~al., 2016, \mn@doi [\aap]
  {10.1051/0004-6361/201525967}, \href
  {https://ui.adsabs.harvard.edu/abs/2016A&A...594A..10P} {594, A10}

\bibitem[\protect\citeauthoryear{{Popescu}, {Yang}, {Tuffs}, {Natale},
  {Rushton}  \& {Aharonian}}{{Popescu} et~al.}{2017}]{Popescu17}
{Popescu} C.~C.,  {Yang} R.,  {Tuffs} R.~J.,  {Natale} G.,  {Rushton} M.,
  {Aharonian} F.,  2017, \mn@doi [\mnras] {10.1093/mnras/stx1282}, \href
  {https://ui.adsabs.harvard.edu/abs/2017MNRAS.470.2539P} {470, 2539}

\bibitem[\protect\citeauthoryear{{Preibisch} et~al.,}{{Preibisch}
  et~al.}{2011}]{Preibisch11}
{Preibisch} T.,  et~al., 2011, \mn@doi [\aap] {10.1051/0004-6361/201116781},
  \href {https://ui.adsabs.harvard.edu/abs/2011A&A...530A..34P} {530, A34}

\bibitem[\protect\citeauthoryear{{Preibisch}, {Flaischlen}, {Gaczkowski},
  {Townsley}  \& {Broos}}{{Preibisch} et~al.}{2017}]{Preibisch17}
{Preibisch} T.,  {Flaischlen} S.,  {Gaczkowski} B.,  {Townsley} L.,   {Broos}
  P.,  2017, \mn@doi [\aap] {10.1051/0004-6361/201730874}, \href
  {https://ui.adsabs.harvard.edu/abs/2017A&A...605A..85P} {605, A85}

\bibitem[\protect\citeauthoryear{{Rebolledo}, {Green}, {Burton}, {Breen}  \&
  {Garay}}{{Rebolledo} et~al.}{2021}]{Rebolledo21}
{Rebolledo} D.,  {Green} A.~J.,  {Burton} M.~G.,  {Breen} S.~L.,   {Garay} G.,
  2021, \mn@doi [\apj] {10.3847/1538-4357/abd7a3}, \href
  {https://ui.adsabs.harvard.edu/abs/2021ApJ...909...93R} {909, 93}

\bibitem[\protect\citeauthoryear{{Reitberger}, {Reimer}, {Reimer}  \&
  {Takahashi}}{{Reitberger} et~al.}{2015}]{Reitberger15}
{Reitberger} K.,  {Reimer} A.,  {Reimer} O.,   {Takahashi} H.,  2015, \mn@doi
  [\aap] {10.1051/0004-6361/201525726}, \href
  {https://ui.adsabs.harvard.edu/abs/2015A&A...577A.100R} {577, A100}

\bibitem[\protect\citeauthoryear{{Seo} et~al.,}{{Seo} et~al.}{2019}]{Seo19}
{Seo} Y.~M.,  et~al., 2019, \mn@doi [\apj] {10.3847/1538-4357/ab2043}, \href
  {https://ui.adsabs.harvard.edu/abs/2019ApJ...878..120S} {878, 120}

\bibitem[\protect\citeauthoryear{{Shull}, {Darling}  \& {Danforth}}{{Shull}
  et~al.}{2021}]{Shull21}
{Shull} M.,  {Darling} J.,   {Danforth} C.,  2021, arXiv e-prints, \href
  {https://ui.adsabs.harvard.edu/abs/2021arXiv210307922S} {p. arXiv:2103.07922}

\bibitem[\protect\citeauthoryear{{Smith}}{{Smith}}{2006}]{Smith06}
{Smith} N.,  2006, \mn@doi [\mnras] {10.1111/j.1365-2966.2006.10007.x}, \href
  {https://ui.adsabs.harvard.edu/abs/2006MNRAS.367..763S} {367, 763}

\bibitem[\protect\citeauthoryear{{Smith}}{{Smith}}{2008}]{Smith08}
{Smith} N.,  2008, \mn@doi [\nat] {10.1038/nature07269}, \href
  {https://ui.adsabs.harvard.edu/abs/2008Natur.455..201S} {455, 201}

\bibitem[\protect\citeauthoryear{{Smith} \& {Brooks}}{{Smith} \&
  {Brooks}}{2007}]{distance}
{Smith} N.,  {Brooks} K.~J.,  2007, \mn@doi [\mnras]
  {10.1111/j.1365-2966.2007.12021.x}, \href
  {https://ui.adsabs.harvard.edu/abs/2007MNRAS.379.1279S} {379, 1279}

\bibitem[\protect\citeauthoryear{{Smith}, {Egan}, {Carey}, {Price}, {Morse}  \&
  {Price}}{{Smith} et~al.}{2000}]{Smith20}
{Smith} N.,  {Egan} M.~P.,  {Carey} S.,  {Price} S.~D.,  {Morse} J.~A.,
  {Price} P.~A.,  2000, \mn@doi [\apjl] {10.1086/312578}, \href
  {https://ui.adsabs.harvard.edu/abs/2000ApJ...532L.145S} {532, L145}

\bibitem[\protect\citeauthoryear{{Sodroski}, {Odegard}, {Arendt}, {Dwek},
  {Weiland}, {Hauser}  \& {Kelsall}}{{Sodroski} et~al.}{1997}]{Sodroski97}
{Sodroski} T.~J.,  {Odegard} N.,  {Arendt} R.~G.,  {Dwek} E.,  {Weiland} J.~L.,
   {Hauser} M.~G.,   {Kelsall} T.,  1997, \mn@doi [\apj] {10.1086/303961},
  \href {https://ui.adsabs.harvard.edu/abs/1997ApJ...480..173S} {480, 173}

\bibitem[\protect\citeauthoryear{{Sun}, {Yang}  \& {Wang}}{{Sun}
  et~al.}{2020a}]{sunRSGC1}
{Sun} X.-N.,  {Yang} R.-Z.,   {Wang} X.-Y.,  2020a, \mn@doi [\mnras]
  {10.1093/mnras/staa947}, \href
  {https://ui.adsabs.harvard.edu/abs/2020MNRAS.494.3405S} {494, 3405}

\bibitem[\protect\citeauthoryear{{Sun}, {Yang}, {Liang}, {Peng}, {Zhang},
  {Wang}  \& {Aharonian}}{{Sun} et~al.}{2020b}]{sunw40}
{Sun} X.-N.,  {Yang} R.-Z.,  {Liang} Y.-F.,  {Peng} F.-K.,  {Zhang} H.-M.,
  {Wang} X.-Y.,   {Aharonian} F.,  2020b, \mn@doi [\aap]
  {10.1051/0004-6361/202037580}, \href
  {https://ui.adsabs.harvard.edu/abs/2020A&A...639A..80S} {639, A80}

\bibitem[\protect\citeauthoryear{{Sun}, {Yang}  \& {Liang}}{{Sun}
  et~al.}{2022}]{sun22}
{Sun} X.-N.,  {Yang} R.-Z.,   {Liang} E.-W.,  2022, \mn@doi [\aap]
  {10.1051/0004-6361/202142394}, \href
  {https://ui.adsabs.harvard.edu/abs/2022A&A...659A..83S} {659, A83}

\bibitem[\protect\citeauthoryear{{Tavani} et~al.,}{{Tavani}
  et~al.}{2009}]{Tavani09}
{Tavani} M.,  et~al., 2009, \mn@doi [\apjl] {10.1088/0004-637X/698/2/L142},
  \href {https://ui.adsabs.harvard.edu/abs/2009ApJ...698L.142T} {698, L142}

\bibitem[\protect\citeauthoryear{{Vall{\'e}e}}{{Vall{\'e}e}}{2014}]{vall14}
{Vall{\'e}e} J.~P.,  2014, \mn@doi [\apjs] {10.1088/0067-0049/215/1/1}, \href
  {https://ui.adsabs.harvard.edu/abs/2014ApJS..215....1V} {215, 1}

\bibitem[\protect\citeauthoryear{{Verner}, {Bruhweiler}  \& {Gull}}{{Verner}
  et~al.}{2005}]{Verner05}
{Verner} E.,  {Bruhweiler} F.,   {Gull} T.,  2005, \mn@doi [\apj]
  {10.1086/429400}, \href
  {https://ui.adsabs.harvard.edu/abs/2005ApJ...624..973V} {624, 973}

\bibitem[\protect\citeauthoryear{{White}, {Breuhaus}, {Konno}, {Ohm}, {Reville}
   \& {Hinton}}{{White} et~al.}{2020}]{White20}
{White} R.,  {Breuhaus} M.,  {Konno} R.,  {Ohm} S.,  {Reville} B.,   {Hinton}
  J.~A.,  2020, \mn@doi [\aap] {10.1051/0004-6361/201937031}, \href
  {https://ui.adsabs.harvard.edu/abs/2020A&A...635A.144W} {635, A144}

\bibitem[\protect\citeauthoryear{{Yang} \& {Aharonian}}{{Yang} \&
  {Aharonian}}{2017}]{Yang17}
{Yang} R.-z.,  {Aharonian} F.,  2017, \mn@doi [\aap]
  {10.1051/0004-6361/201630213}, \href
  {https://ui.adsabs.harvard.edu/abs/2017A%26A...600A.107Y} {600, A107}

\bibitem[\protect\citeauthoryear{{Yang} \& {Liu}}{{Yang} \&
  {Liu}}{2022}]{Yang22}
{Yang} R.-Z.,  {Liu} B.,  2022, \mn@doi [Science China Physics, Mechanics, and
  Astronomy] {10.1007/s11433-021-1777-4}, \href
  {https://ui.adsabs.harvard.edu/abs/2022SCPMA..6519511Y} {65, 219511}

\bibitem[\protect\citeauthoryear{{Yang}, {de O{\~n}a Wilhelmi}  \&
  {Aharonian}}{{Yang} et~al.}{2018}]{Yang18}
{Yang} R.-z.,  {de O{\~n}a Wilhelmi} E.,   {Aharonian} F.,  2018, \mn@doi
  [\aap] {10.1051/0004-6361/201732045}, \href
  {https://ui.adsabs.harvard.edu/abs/2018A%26A...611A..77Y} {611, A77}

\bibitem[\protect\citeauthoryear{{Zabalza}}{{Zabalza}}{2015}]{Zabalza15}
{Zabalza} V.,  2015, in 34th International Cosmic Ray Conference (ICRC2015).
  p.~922 (\mn@eprint {arXiv} {1509.03319})

\makeatother
\end{thebibliography}

% Don't change these lines
\bsp	% typesetting comment
\label{lastpage}
\end{document}